\documentclass[fleqn,usenatbib]{mnras}

\usepackage{newtxtext,newtxmath}
\usepackage[T1]{fontenc}
\usepackage{ae,aecompl}

\usepackage{graphicx}	% Including figure files
\usepackage{amsmath}	% Advanced maths commands
\usepackage{amssymb}	% Extra maths symbols
\usepackage{multicol}   % Plot figures side by side
\usepackage[colorinlistoftodos]{todonotes} % Editing notes

\newcommand{\simgt}{\lower 2pt \hbox{$\, \buildrel {\scriptstyle >}\over {\scriptstyle\sim}\,$}}
\newcommand{\simlt}{\lower 2pt \hbox{$\, \buildrel {\scriptstyle <}\over {\scriptstyle\sim}\,$}}

\title[Deficit of Red Galaxies in Cosmic Voids]{Deficit of Luminous and Normal Red Galaxies in Cosmic Voids}

\author[Bruton et al.]{
Sean T. Bruton,$^{1}$\thanks{E-mail: seanbruton@ou.edu}
Xinyu Dai,$^{1}$
Eduardo Guerras$^{1}$
Ferah A.\ Munshi$^{1,2}$
\\
$^{1}$Homer L. Dodge Deparment of Physics and Astronomy, University of Oklahoma, Norman 73019\\
$^{2}$Department of Physics \& Astronomy, 6301 Stevenson Center, Vanderbilt University, Nashville, TN 37235\\
}

\date{Accepted XXX. Received YYY; in original form ZZZ}

\pubyear{2019}
\begin{document}
\label{firstpage}
\pagerange{\pageref{firstpage}--\pageref{lastpage}}
\maketitle

\begin{abstract}
We construct a sample of 10,680 wall galaxies and 3,064 void galaxies with $M_R \simlt -20$ by cross referencing a void catalog from literature with Baryon Oscillation Spectroscopic Survey (BOSS) CMASS and WiggleZ galaxies, where the CMASS survey targets redder galaxies and the WiggleZ survey targets bluer galaxies.  Comparing the density profiles of the red and blue galaxies as a function of the void radius, we find that the number ratio of red-to-blue galaxies increases with distances from the void centers, suggesting a deficit of luminous and normal red galaxies in voids. We find a mean (g -- r) magnitude color of 1.298 and 1.210 for the wall and void galaxies, respectively, when considering the combined red and blue sample, which is found to be a significant difference. However, when considering the blue and red samples separately, we find no significant color difference. We conclude that the constituents galaxies of each population, rather than intrinsic color difference, is the main driver in the apparent average color difference of galaxies in voids and walls, indicating a deficit of luminous and normal red galaxies in voids. Our analysis suggests that the primary environmental-dependence effect on galaxy evolution for normal and luminous galaxies between void and wall regions is manifested in the number of red galaxies, which depends on the environmental-dependent merger history.  Using a semi-analytic simulation model, we can successfully reproduce the apparent color difference between the void and wall galaxies.
\end{abstract}


%%%%%%%%%%%%%%%%%%%%%%%%%%%%%%%%%%%%%%%%%%%%%%%%%%

%%%%%%%%%%%%%%%%% BODY OF PAPER %%%%%%%%%%%%%%%%%%

\section{Introduction}
On the largest scale, the universe is homogeneous and isotropic, but on scales up to 100~Mpc, galaxies clump together into filaments, walls, and voids to form a sponge-like foam, the large scale structure \citep{bharadwaj04}. Large, under-dense regions of space which exist between the filaments, clusters, and walls of galaxies are known as cosmic voids and typically have radii $30 h^{-1}Mpc < r < 80 h^{-1}Mpc$ \citep{mao2017}. These cosmic voids make up about 60\% of the the volume of the universe, while they contain only about 7\% of the galaxies \citep{pan12}. The galaxies which occupy these voids are known as void galaxies. Galaxies are thought to form and evolve from both the collapse of dark matter halos as well as mergers of halos, with the additional gas following more complicated gas physics attached on the backbone of the dark matter.
The under-dense void environments provide an ideal environment for studying galaxy formation and evolution, as gravitational interactions with neighbors are less frequent, and it is possible to separate the role of mergers in these processes. Thus, void galaxies can provide insight on the environmental-dependence of galaxy formation/evolution with the minimal impact from mergers. However, not all void galaxies are isolated -- filaments within voids permit gravitational interactions even among void galaxies. For example, \citet{kreckel2012} found that 5 of the 60 Void Galaxy Survey (VGS) galaxies are interacting with a companion. Following up on this, \citet{beygu2013} looked in depth at a filament of three galaxies in the VGS.

With such promise, void galaxies have been a prominent area of research in the previous years. It has been found that void galaxies are typically blue, late-type, star-forming galaxies \citep{szomoru1996,popescu1997,grogin1999, rojas2003, rojas2004, sorrentino2006, tavasoli2015}. While this leads to a clear difference across the void and wall galaxies as a whole, a search for quantitative differences in the properties of void and wall galaxies has been more evasive. 
Both \citet{croton2005} and \citet{hoyle2005} found no significant difference between the faint-end slopes in the luminosity functions between void and wall galaxies, with the qualification that \citet{croton2005} were considering the full 2dFGRS galaxy sample. However, it is noteworthy that \citet{croton2005} find that the faint-end luminosity function of early type galaxies steepens with increasing density. In contrast, \citet{park2007} found significant difference between void and wall galaxies at the faint-end slopes. These results could lead to contradicting conclusions about the population of dwarf galaxy populations in voids. In terms of differences in color between wall and void galaxies, \citet{rojas2004} find that void galaxies are bluer than wall galaxies, and that this behavior is only partially explained with the absence of luminous red galaxies in voids. \citet{patiri2006} found evidence that wall and void galaxies do not differ in their individual properties, particularly with regards to color distribution, bulge-to-total ratios, and the specific star formation rate when accounting for color. More recently, \citet{hoyle2012} found that late-type void galaxies seem to be bluer than late-type wall galaxies, and the same could be said for their early-type galaxies. In this study, we will focus on a statistical analysis of a large sample of wall and void galaxies, and as such will be washing out the minor effects of merging void galaxies or well-isolated wall galaxies, which would cross-contaminate the samples to a small degree.

Our sample in this study is unique in its combination of two distinct spectroscopic surveys targeting red and blue galaxies, which will facilitate a quantitative study on the color differences and red-to-blue galaxy number ratio between wall and void galaxies, especially for the luminous and normal ones.
These measurements can be compared with numerical or semi-analytical simulations to test the environmental-dependence galaxy formation models. 
To the end of studying these void galaxies, we use photometry from the Sloan Digital Sky Survey (SDSS) CMASS sample \citep{york2000,alam2015}, WiggleZ Dark Energy Sky Survey \citep{drinkwater2010}, and a void catalog \citep{mao2017} to identify and study void galaxies. The SDSS CMASS sample covers about 10,000 square degrees, while the WiggleZ covers about 1,000 square degrees. The majority of the WiggleZ sample overlaps with the CMASS sample, as shown in \citep{beutler2016}. 

In \S2, we discuss the two galaxy surveys as well as the void catalog. In \S3, we present our methodology. In \S4 and 5, we present our analysis results and comparisons with semi-analytic model predictions.  Discussion is given in \S6.  We assume a flat cosmology with $\rm H_0=70~km~s^{-1}~Mpc^{-1}$, $\rm \Omega_M=0.27$, and $\rm \Omega_\Lambda=0.73$ throughout the paper.

\section{Datasets}

\begin{figure*}
\begin{multicols}{2}
    \includegraphics[width=\linewidth]{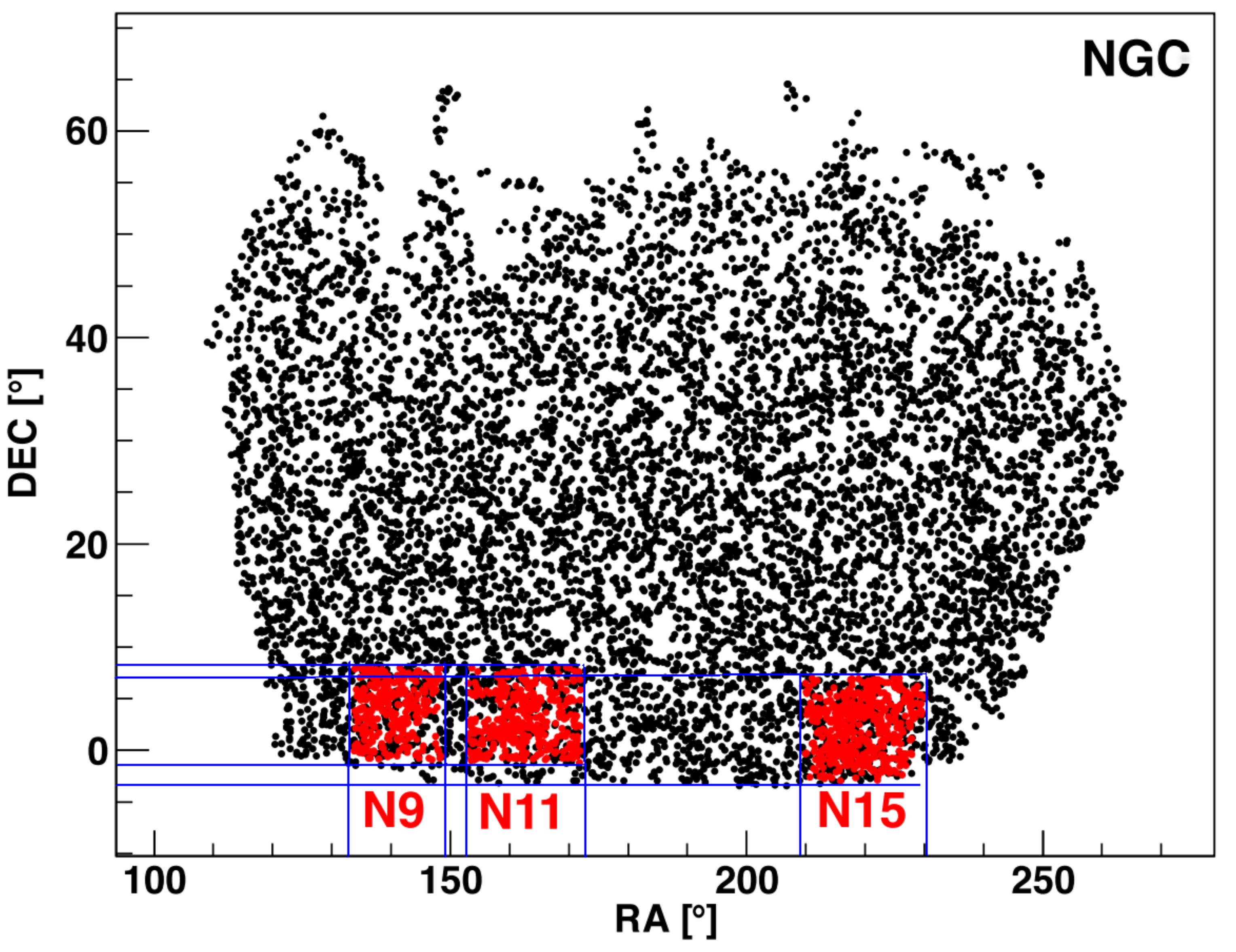}\par 
    \includegraphics[width=\linewidth]{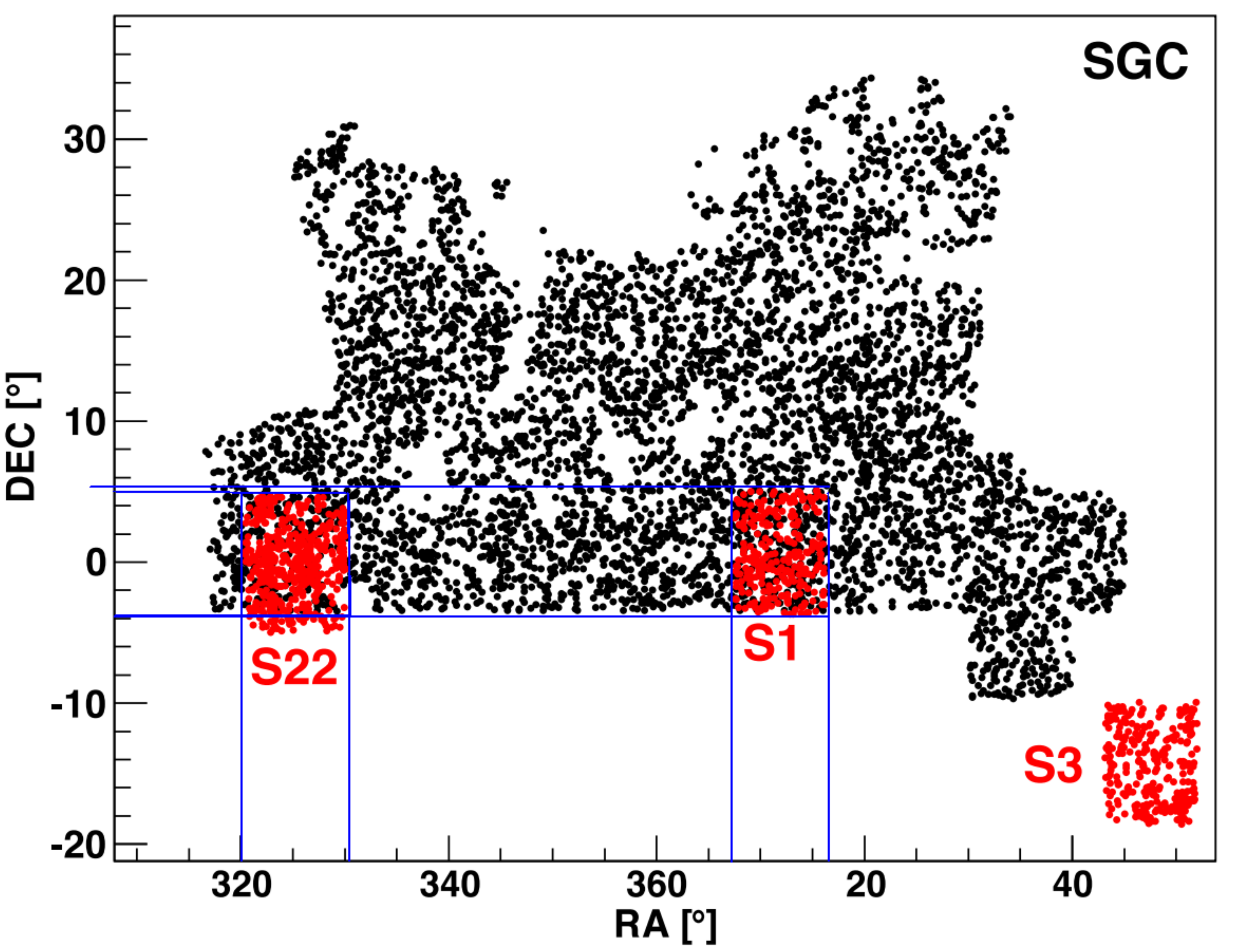}\par 
    \end{multicols}
\caption{Schematic plot of the sky coverage of CMASS and WiggleZ surveys modified from \citet{beutler2016}, where we added the blue lines to facilitate identifying the approximate overlapping sky coverage between the CMASS  (black) and WiggleZ (red) surveys. The left frame shows the North Galactic Cap (NGC) overlap regions, while the right frame shows the South Galactic Cap (SGC.) Only a random 3\% of all galaxies are plotted.}\label{fig:1}
\end{figure*}

\begin{figure*}
\begin{multicols}{2}
    \includegraphics[width=\linewidth]{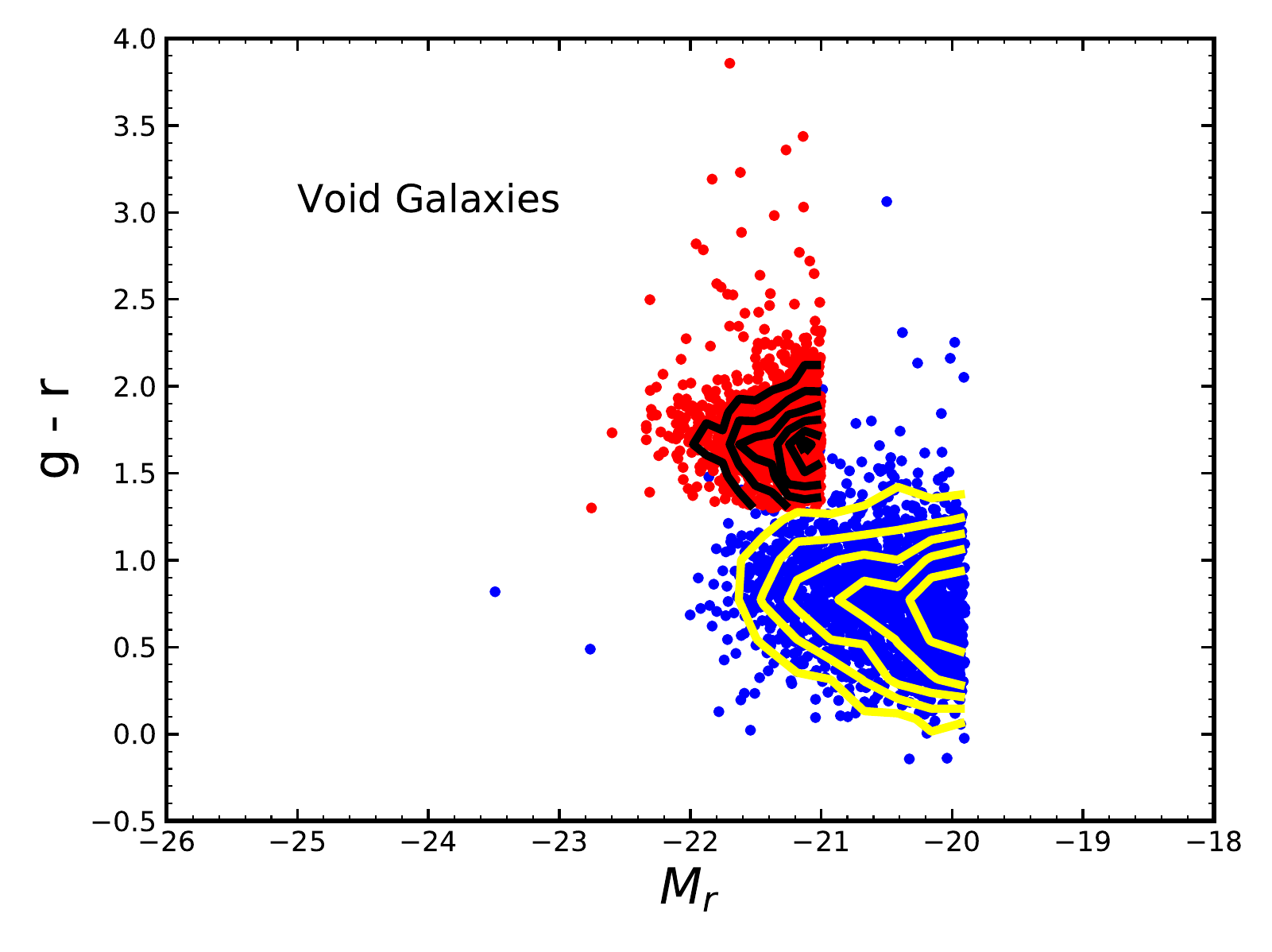}\par 
    \includegraphics[width=\linewidth]{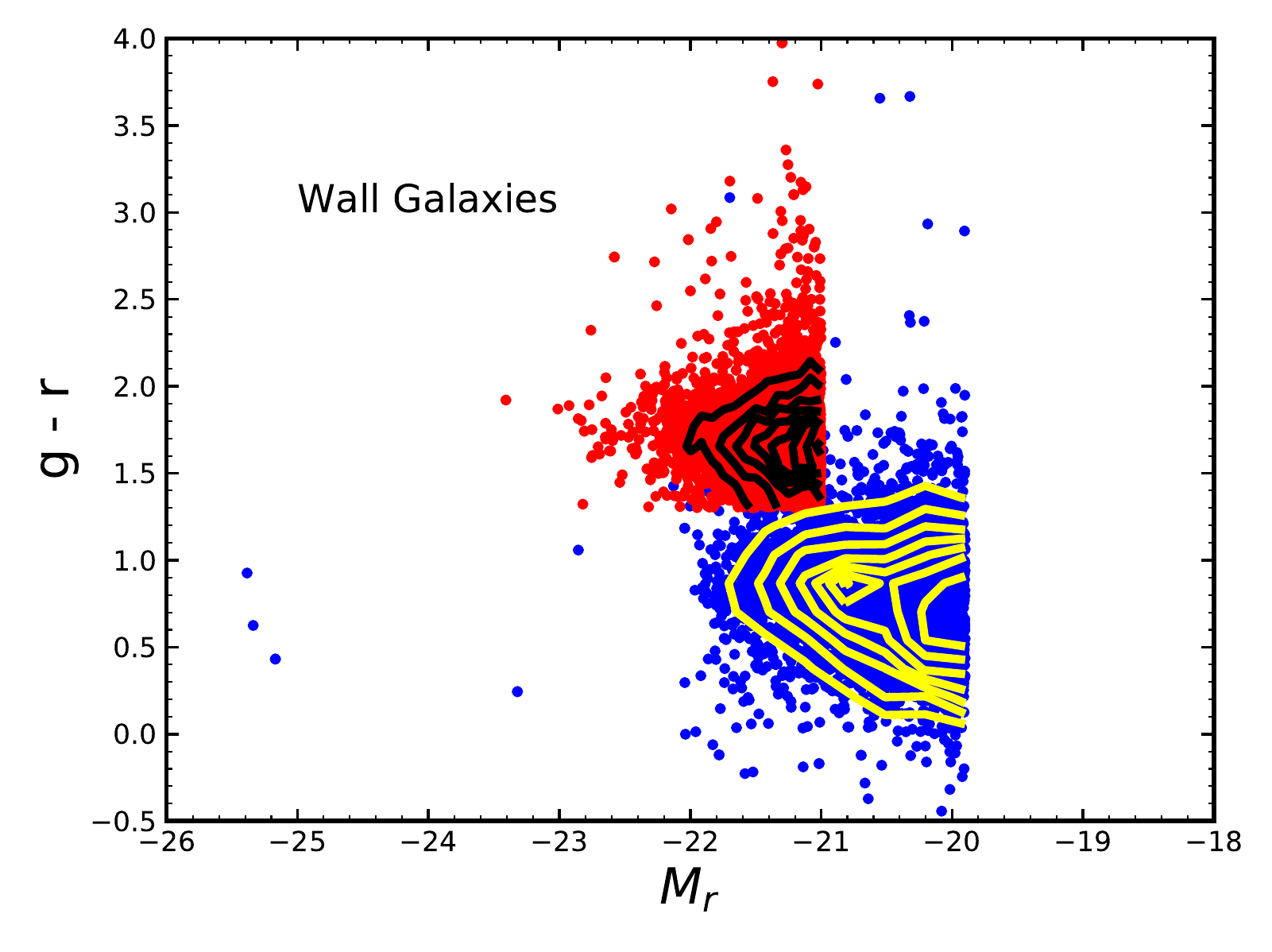}\par 
    \end{multicols}
\caption{Color Magnitude diagrams for both the wall (left) and void (right) galaxies, where we plot the $g - r$ color vs the absolute magnitude in the SDSS $r$ filter. We apply the absolute magnitude cuts of $M_R < -21.0$ to the CMASS sample and $M_R < -19.9$ to the WiggleZ sample. The CMASS and WiggleZ magnitude cut values are chosen to correspond with the peak of the magnitude histograms, as shown in Figure \ref{fig:3} The results are shown to be insensitive to the exact CMASS or WiggleZ magnitude cut value. The CMASS galaxies are plotted in red, while the WiggleZ galaxies are plotted in blue. Black and yellow contour lines help to identify the red sequence and blue cloud. It is apparent that the CMASS galaxies mainly occupy the red sequence, while the WiggleZ galaxies mainly occupy the blue cloud.}\label{fig:2}
\end{figure*}

\subsection{The CMASS Sample}
The CMASS sample is part of the Baryon Oscillation Spectroscopic Survey (BOSS), conducted with the 2.5 meter telescope at the Apache Point Observatory in New Mexico. BOSS covered 10,000\,deg$^2$ of sky, observing about 1.5 million galaxies as faint as $i = 19.9$ mag at a redshift range  $0.15 < z < 0.7$. BOSS is split into two samples --- the LOWZ sample from $0.15 < z < 0.43$ and the CMASS sample from $0.43 < z < 0.7$. For our analysis, we use only the CMASS sample. The LOWZ sample has fewer voids overall, and after applying the sample selections discussed in \S3 only 22 voids remain. This results in a much smaller sample of both void and wall galaxies, only 1378 galaxies from the LOWZ sample before any additional data screens. The small number statistics  makes it difficult to interpret the analysis results, and thus, we work only with the CMASS sample. The CMASS sample favors red galaxies, although to a lesser extent than the LOWZ sample, but both are not devoid of blue galaxies. Galaxy light is collected in the SDSS five band system --- \textit{u}, \textit{g}, \textit{r}, \textit{i}, and \textit{z}, centered at 3551\AA, 4686\AA, 6165\AA, 7481\AA, and 8931\AA, respectively. We use the \textit{g} and \textit{r} band magnitudes and the spectroscopic redshift in this analysis.

\subsection{The WiggleZ Survey}
The WiggleZ Dark Energy Survey was conducted at the Anglo-Australian telescope, covering 1,000\,deg$^2$ of sky. We work with the first public data release, which contains 81,362 galaxies. The WiggleZ Survey was designed to measure the Baryon Acoustic Oscillation (BAO) using blue galaxies with strong emission lines originating from star-forming regions. These strong emission lines allow quick and accurate detection of redshifts, facilitating redshift distance measurements which are necessary for the measurement of the BAO signal. The target galaxies were selected using GALEX satellite data. 90\% of the target galaxies fall within redshift 0.2 to 1.0. A sky-position target match is done between the GALEX data and SDSS data, matching with a confidence level of 95\%. Thus, we have access to the SDSS optical data for these galaxies, and once again use the $g$ and $r$ magnitudes and the spectroscopic redshift in this analysis. 

\subsection{The CMASS-WiggleZ Overlapping Region}
The majority of the WiggleZ sky coverage is within the CMASS sky coverage, resulting in about 560\,deg$^2$ of overlap \citep{marin2016}. The CMASS and WiggleZ surveys overlap at a redshift of $0.43 < z < 0.70$, limited by the CMASS sample. This large area of overlap gives us a sizable population of galaxies to work with, as we shall see in section \S3.

\subsection{Incompleteness of the Samples}
For the WiggleZ sample, the incompleteness issue is less important due to the targeting strategy \citep{blake2010}.
Although the CMASS sample suffers from various incompleteness factors, such as the redshift failure, fiber collision, and other systematic effects \citep[e.g.,][]{beutler2016}, the analysis in this paper focus on the relative comparisons between the void and wall galaxies.  Therefore, these incompleteness factors should not affect our result significantly, unless a factor has different effects for the void and wall regions.  We assume that the redshift failure and other systematic factors behave similarly in the void and wall regions.
The fiber collision factor could affect the void and wall directions differently, because there are more galaxies at the wall directions that can result in a higher likelihood of fiber collision. 
However, for the redshift range considered $0.43<z<0.70$, the fiber collision beam corresponds to 0.55--0.9~Mpc, while our voids are up to 60 Mpc in radius, much larger than the fiber collision sizes and galaxies are sparsely populated in voids. 
Thus, one may presume that the fiber collision factor will have minimal affects on our galaxy counts between the voids and walls. 
In addition, voids are three-dimensional, and there are foreground and background galaxies in their directions. Therefore, the two-dimensional surface density contrast may not be as large in the void directions when compared to the wall directions, resulting in similar fiber collision rates. 
To check the fiber collision effect more carefully, we can calculate the nearest neighbor distance for the galaxies in the CMASS and WiggleZ samples, using WiggleZ sample as a reference because it suffers less from imcompleteness issues. 
Here, we count the total number of galaxies which have a neighbor within 1~Mpc. For the CMASS sample, we find that 59 of the final sample of 7,092 galaxies (see \S3 for the selection cuts) have a nearest neighbor within 1Mpc, or 0.8\%. In the WiggleZ sample, 44 of the 6,652 galaxies have a nearest neighbor within 1Mpc, or 0.6\%. If fiber collisions were significant in CMASS sample, we would expect to see a significant difference between the number of galaxies with a neighbor within 1~Mpc across the two samples.
We conclude that the CMASS incompleteness factors do not affect the main results of our paper.

\subsection{The Void Catalog}
The void catalog we use in our study was constructed by \citet{mao2017} using BOSS data, with 7444 and 3199  voids in CMASS and LOWZ samples, respectively. We use only the 7744 CMASS voids in our work. Voids are found using the ZOBOV algorithm, which performs Voronoi tessellation to segment the volume of the survey among the galaxy population \citep{neyrinck2008}. A watershed transformation is then applied to group Voronoi cells together based on local minimum densities. This results in many voids which may have arbitrary shapes, not limited to spherical shapes nor conjoined spherical shapes. 
Although the voids are not necessarily spherical, an effective radius is attributed to each. This co-moving radius is calculated by summing the Voronoi volumes of each galaxy in a void, then using
\begin{equation}\label{eq:1}
R_{eff} \equiv (\frac{3}{4\pi}V)^{1/3}.
\end{equation}
The voids have an average radius of $\simeq60$ Mpc.
We use the void sky positions along with their effective radius to determine if a CMASS or WiggleZ galaxy is a void galaxy in the three-dimensional space, utilizing the accurate spectroscopic redshifts for distance measurements from both samples. We exclude voids which have greater than a 30\% chance of being the result of Poisson noise. This leaves us with 2404 north voids and 849 south voids, for a total of 3253 voids with less than 30\% chance of being the result of noise.

\section{Sample Selection}
\subsection{Classification of Wall and Void Galaxies}
We begin with identifying the void galaxies in the CMASS and WiggleZ overlapping regions. The first step is to make a cut of the CMASS sample to include only sky coordinates which overlap with the WiggleZ sample, as shown in Figure \ref{fig:1} \citep[modified from][]{beutler2016}. The values for these estimates are taken from \citet{marin2016}. By excluding CMASS galaxies outside the overlap region, we can ensure that our results are not the product of survey boundary differences between the WiggleZ and CMASS samples. 

As mentioned previously, each void has a given effective radius, $R_{eff}$. We utilize this effective radius in selecting our sample of galaxies. If a galaxy falls within $1.5*R_{eff}$ of a given void, it is added to our catalog of galaxies. At this phase, we work with voids smaller than 60 Mpc, which coincides with the peak of the approximately Gaussian distribution of voids' effective radii. 
When calculating the co-moving distance, d, between the each galaxy and every void, we use a flat $\Lambda$-CDM with $\Omega_M = 0.27$. 
Galaxies within one $R_{eff}$ are deemed void galaxies, while galaxies within 1 $R_{eff}$ < d < 1.5 $R_{eff}$ are deemed wall galaxies. While this definition of void and wall galaxies will inevitably have contaminations across populations, this is all but unavoidable due to the voids' complex shapes. 
Using the large void catalog, the difference in the void shape will average out in the stacked void profiles.

To further ensure that the voids we are sampling represent overlap regions between the two surveys, we impose the condition that a void and its shell must contribute at least five galaxies from both the CMASS and WiggleZ survey for the contained void galaxies to be included in our catalog. We are left with a catalog of 13,928 CMASS galaxies and 8,881 WiggleZ galaxies across 188 voids.

\subsection{Additional Data Screenings}
\begin{figure*}
\begin{multicols}{2}
    \includegraphics[width=\linewidth]{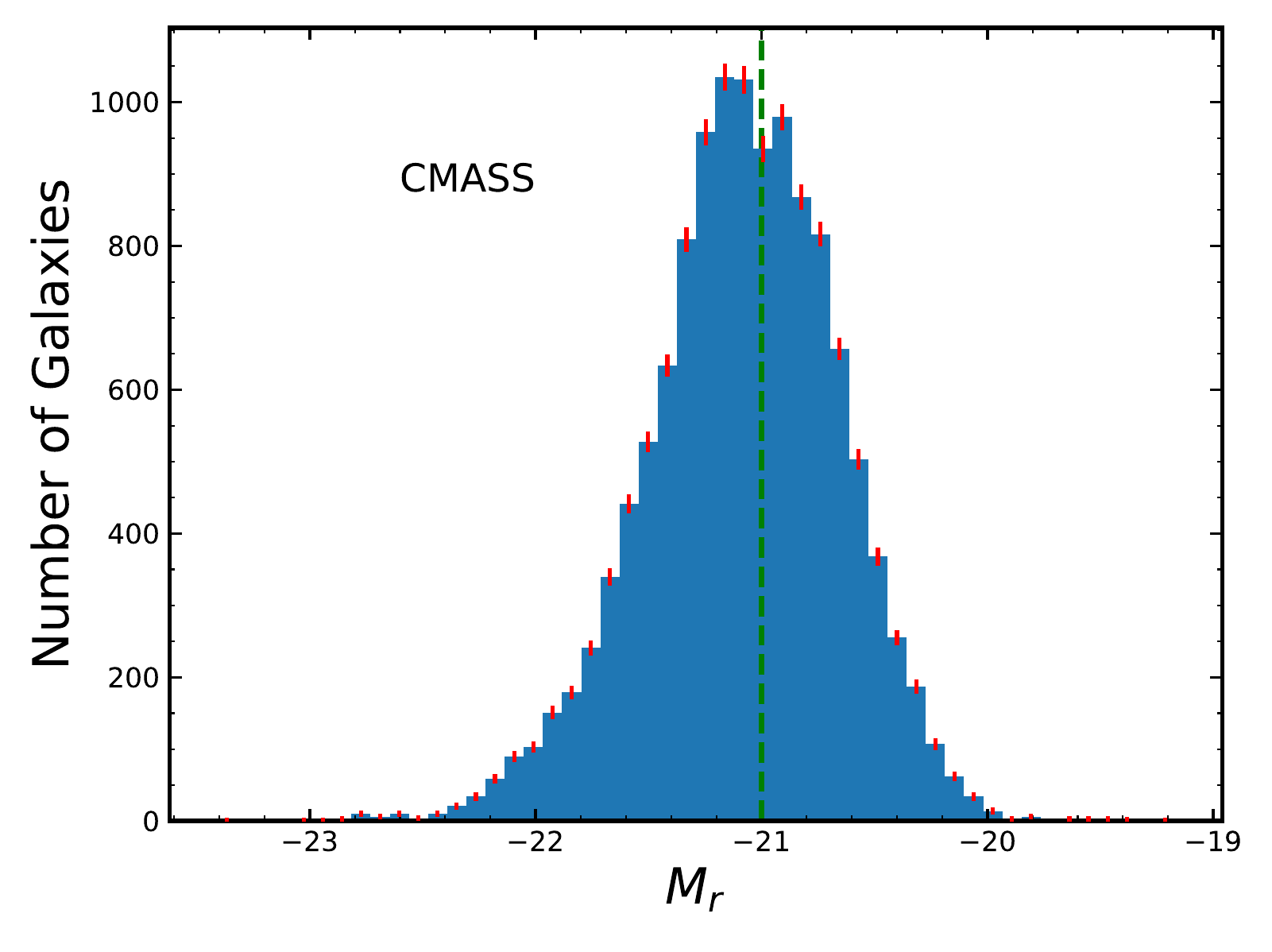}\par 
    \includegraphics[width=\linewidth]{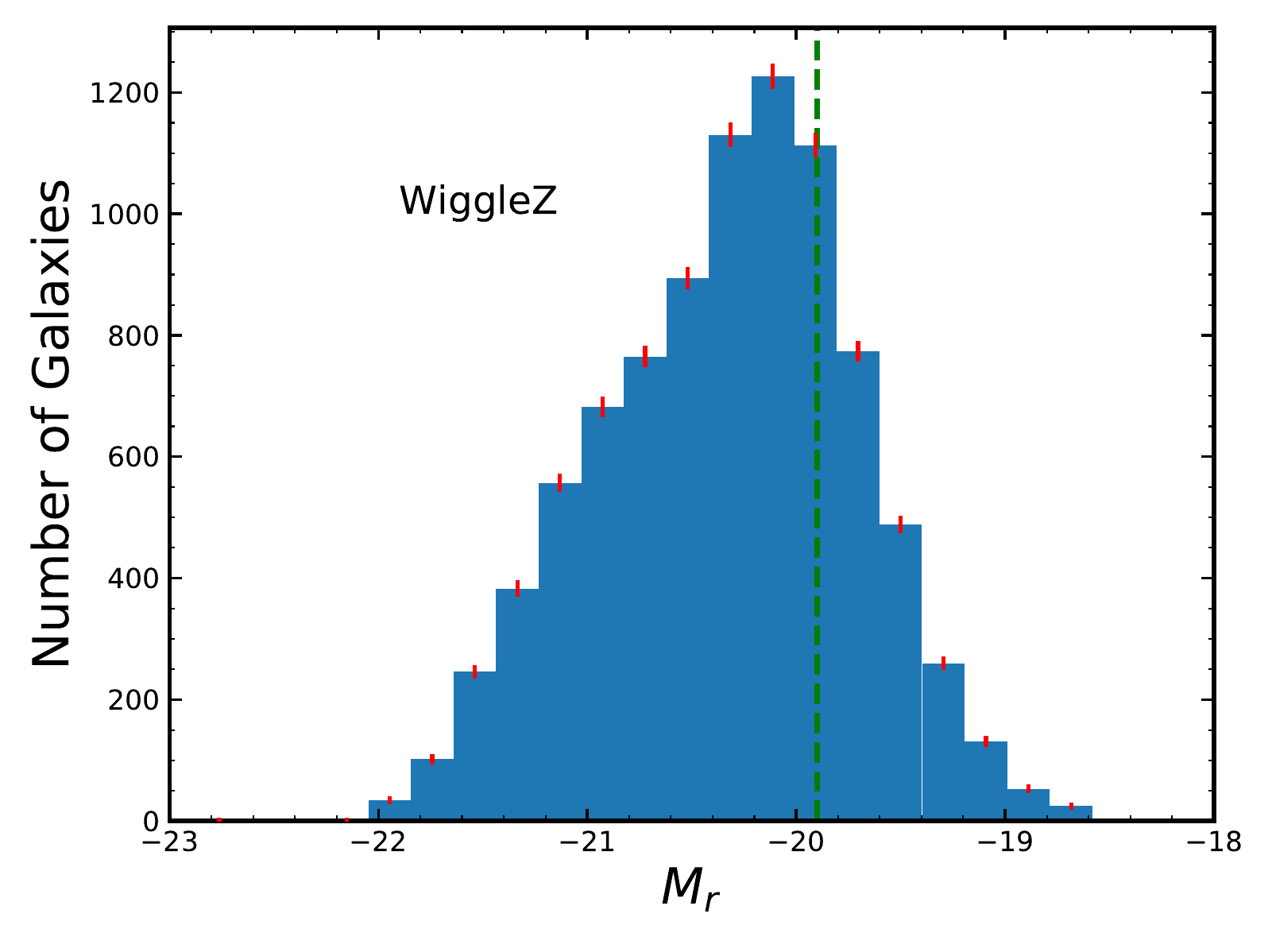}\par 
    \end{multicols}
\caption{The magnitude distributions of galaxies in the CMASS (left) and WiggleZ (right) populations. Error bars are given by $\delta = \sqrt{N}$ and shown in red. The chosen magnitude cuts are shown as the green dashed line, eliminating selection bias of faint galaxies.}\label{fig:3}
\end{figure*}

To further ensure the completeness of the sample, We select only galaxies more luminous than $M_R < -21.0$ and $M_R < -19.9$ for CMASS and WiggleZ samples, respectively. 
Here, the absolute magnitudes are calculated based on the SDSS apparent magnitudes, our assumed cosmology, and the K-corrections from \citet{assef2010}.
These absolute magnitude cuts essentially limit our study to only luminous galaxies.
The WiggleZ and CMASS magnitude cuts are both chosen to correspond with the peak of the galaxies' magnitude distribution as shown in Figure \ref{fig:3}. These restrictions are applied to eliminate bias from including more faint nearby galaxies. 

\begin{figure}
    \includegraphics[width=\columnwidth]{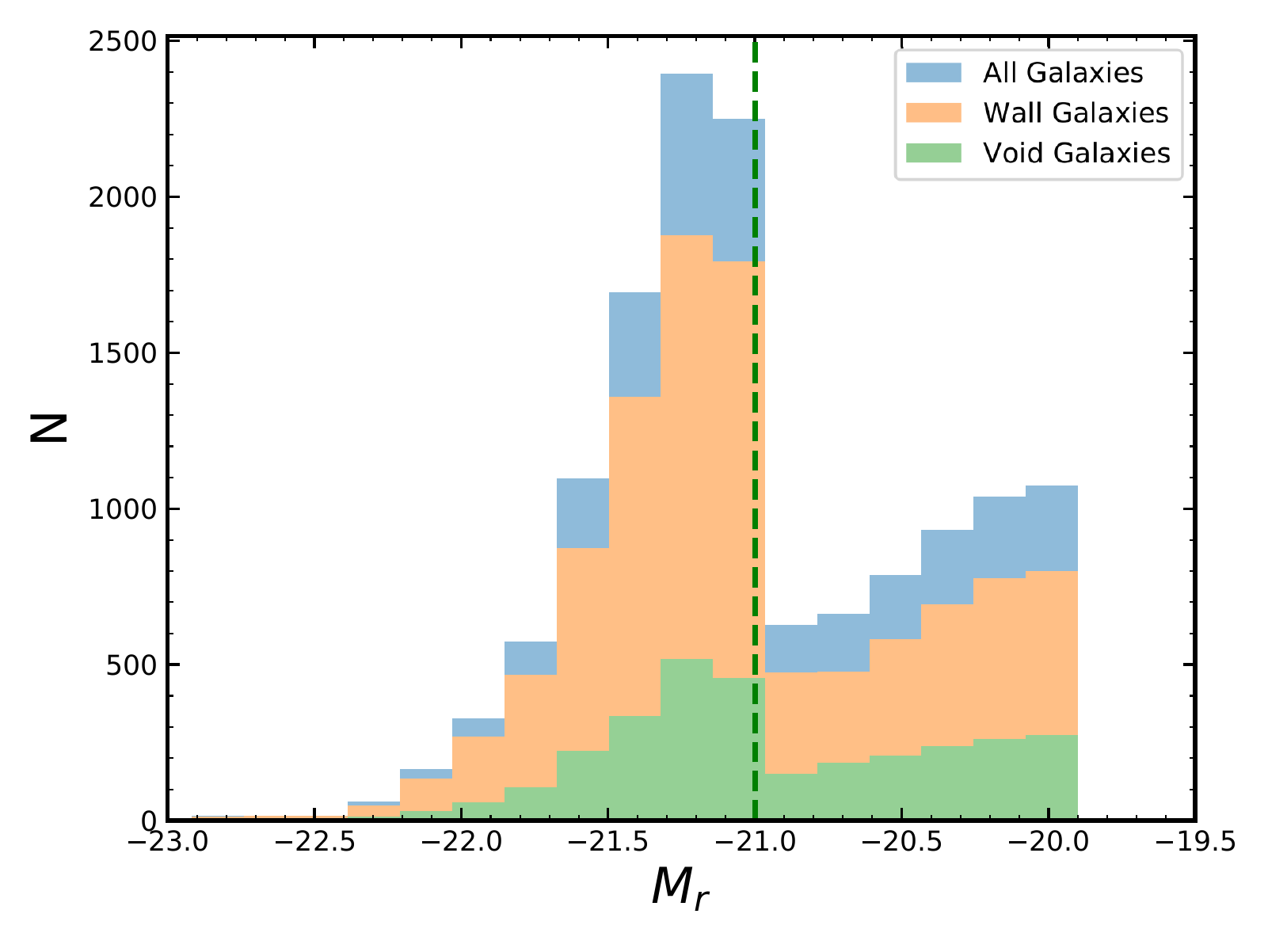}
    \caption{The absolute magnitude distributions of all galaxies, wall galaxies, and void galaxies. The dashed green line show the faint magnitude limit of the CMASS sample, which is the reason for the jump in the distribution at magnitudes brighter than $M_R \simeq -21.0.$. Generally, the distributions of wall and void galaxies follow similar shapes.}\label{fig:all_mag_distr}
\end{figure}

Figure \ref{fig:all_mag_distr} shows the distribution of absolute magnitudes for all galaxies, wall galaxies, and void galaxies. We see that, generally, the wall and void populations follow similar distributions in absolute magnitude. The dashed green line is the dim magntidue cut for the CMASS sample. The onset of the CMASS sample is the reason for the large jump in the distribution at $M_r \simeq -21.0$.

We have tested selecting different magnitude cuts for the CMASS and WiggleZ samples and find that our results are not dependent on the chosen values. For example, applying a cut at $-21.0$ for the WiggleZ sample, as the CMASS sample, is more restrictive and increases the noise, but our fundamental findings do not change. Applying a magnitude cut at $-19.9$ for the CMASS galaxies simply decreases the noise at the expense of using less luminous galaxies with less well-defined selections. We have also tested applying bright magnitude cuts to both the CMASS and WiggleZ samples. Again, we find that none of our findings change with the bright magnitude selection for either the CMASS or WiggleZ samples. Of particular interest, when the two samples have precisely the same faint and bright magnitude cuts, none of our results change. The resulting wall and void galaxy populations always have a significantly different mean $M_r$ magnitude, even when the magnitude limits are the same on both populations. This seems to be because void galaxies tend to be less luminous. 
A dedicated modeling of the luminosity function is needed to further confirm this difference in the mean $M_r$ magnitude between void and wall galaxies.

We also impose a color cut on the CMASS sample: the $g - r$ color of the CMASS sample must be $(g - r)> 1.3$. This eliminates the bluest of galaxies from the CMASS sample, leaving the CMASS sample as the redder galaxies and the WiggleZ sample as the bluer galaxies. This facilitates a ratio comparison, which we will discuss in \S4. These cuts are evident in Figure \ref{fig:2}, which shows the galaxies' color-magnitude diagram between the $(g - r)$ color and the $M_r$ magnitude. Color-magnitude diagrams of galaxies have been shown to cluster in two distinct regions: the red sequence and the blue cloud, which are both apparent. The red sequence is composed primarily of the CMASS galaxies, and the blue cloud is composed primarily of the WiggleZ galaxies. Although void and wall galaxies occupy both the red sequence and the blue cloud, we can make more qualitative statements using statistical testing, as we will show in \S4.

After these cuts, we are left with 7,092 CMASS galaxies and 6,652 WiggleZ galaxies. This is 51\% of the original CMASS sample and 73\% of the original WiggleZ sample. Of the 7,092 CMASS galaxies, 5729 are wall galaxies while 1363 are void galaxies. For comparison, the WiggleZ population has 4951 wall galaxies and 1701 void galaxies. While our void sample is not as large as the ~88,000 void galaxies studied in \citet{hoyle2012}, it is larger than many other previous studies which have studied $\sim1000$ void galaxies. Further, by including the WiggleZ sample and detecting a marked increase in density as we move out of the void, we are able to validate the voids from \citet{mao2017} with an independent sample. It is also noteworthy that the voids used here are generally larger than those used in \citet{hoyle2012}, which are based on the void catalog from \citet{pan12}.

We are now equipped with a large catalog of blue and red void galaxies in the overlap region of CMASS and WiggleZ surveys. 

Finally, we note that while we exclude voids with radius larger than 60 Mpc, all of our results are robust against including voids of larger radius.

\section{Stacking Results}
We use our combined catalog from CMASS and WiggleZ galaxies in the overlapping regions of the two surveys to complete the rest of our analysis. 
We first check the stacked galaxy density profile of the voids using two methods, stacking by the physical or effective radius units.
We expect that the stacked galaxy density would increase as a function of distance from the void center.

\begin{figure*}
\begin{multicols}{2}
    \includegraphics[width=\linewidth]{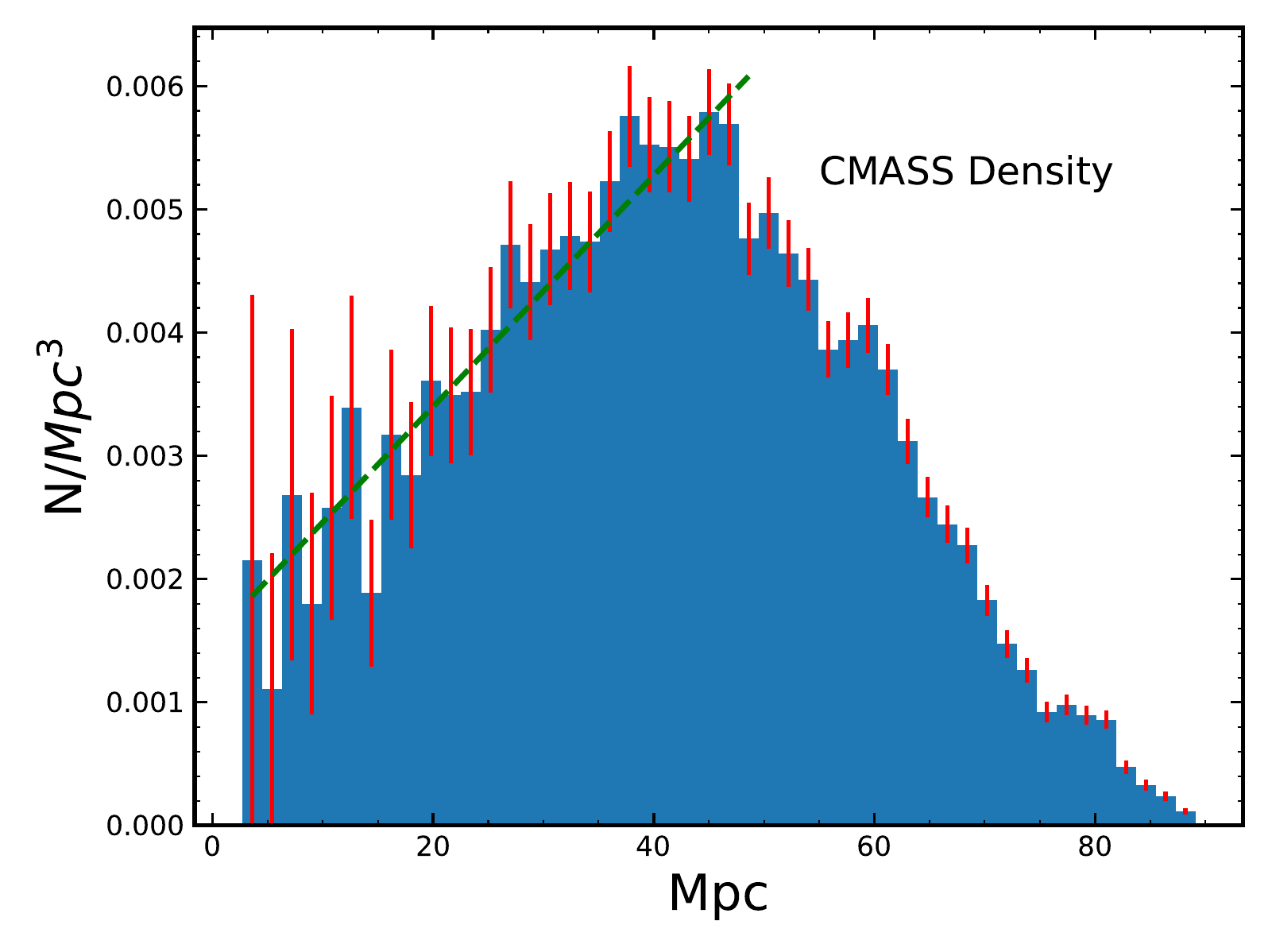}\par 
    \includegraphics[width=\linewidth]{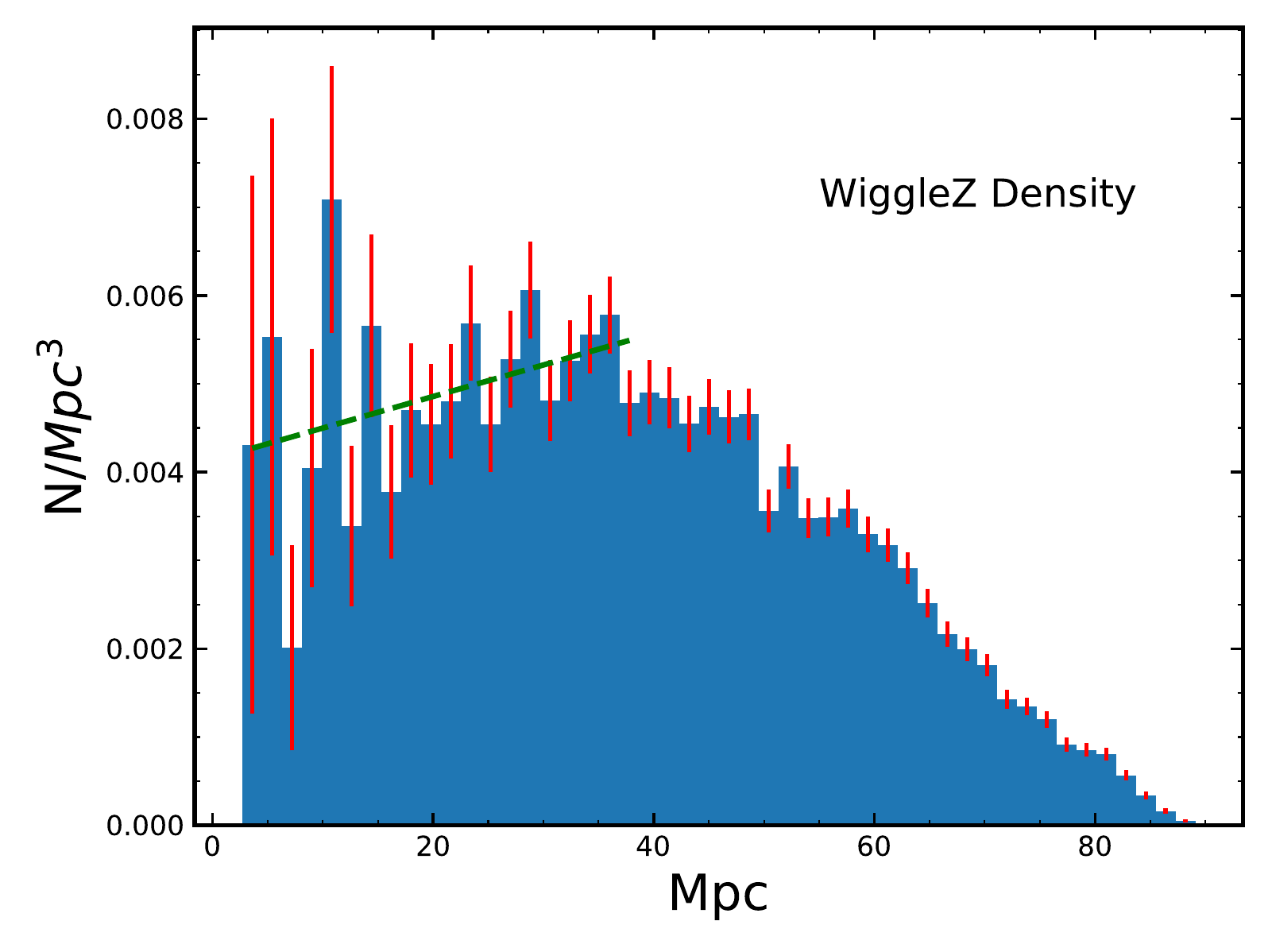}\par 
    \end{multicols}
\caption{Galaxy number density as a function of void radius with the uncertainties shown in red. The left plot shows the CMASS sample, and the right shows the WiggleZ sample. Both plots generally show an increase in density up to 40--60 Mpc, close to our maximum void selection radius, then tamper off. Although the WiggleZ data is noisier, the increasing trend looks plausible though visual examination. Shown with the green dashed line, we fit both density profiles with a linear fit, given by equation \ref{eq:2}. This fit is up to 50 Mpc for the CMASS sample and 40 Mpc for the WiggleZ sample, both corresponding to the maximum of the histograms. The CMASS fit deviates from zero by 13$\sigma$, while the WiggleZ slope deviates by 1.6$\sigma$.}\label{fig:4}
\end{figure*}

Figure \ref{fig:4} shows the stacked density of galaxies as a function of distance from the void center in the physical unit (Mpc) with a bin size of 1.8 Mpc.
In each bin, we co-add the number of galaxies in each void of that bin then divide the co-moving volume of the shell with the Poission uncertainties assumed, 
\begin{equation}\label{eq:3}
\delta = \frac{\sqrt{N}}{\frac{4}{3}\pi(r_2^3-r_1^3)}\,
\end{equation}
where $N$ is the number of galaxies in each bin and $r_2$ and $r_1$ denote the outer and inner edges of the bin, respectively. When stacking on the physical scale, we combine each of the voids using their own respective size, keeping galaxies out to $1.5*R_{eff}$.
Here, we see the galaxy density increase as we move out from the void centre up to 40--50\,Mpc  then fall off as we hit our void size limit, a result of stacking voids of various sizes up to 60 Mpc. 
We apply a linear fit to quantify the stacked void profiles, out to 50 and 40\, Mpc for CMASS and WiggleZ samples respectively, 
given by 
\begin{equation}\label{eq:2}
Ratio = 10^{-3} A \frac{r}{50 Mpc} + B.  
\end{equation}
The CMASS density profile yields a slope of 4.69 $\pm$ 0.37 to 50 Mpc, corresponding to the distance where the galaxy density begins to fall off due to the physical unit limitation. 
This gives a 13$\sigma$ deviation from the zero slope, confirming that we have stacked truly void regions.
The WiggleZ density profile has a slope of 1.8 $\pm$ 1.1 to 40 Mpc, corresponding to only 1.6$\sigma$ deviation from zero. 
Although the WiggleZ void profile is not as significant as the CMASS profile, this is mainly due to the large uncertainties of the void profile near the centre, where galaxy numbers are divided by small shell volumes (Figure~\ref{fig:4}).
This is especially limiting for the WiggleZ sample due to its relatively small number of galaxies. 

\begin{figure*}
\begin{multicols}{2}
    \includegraphics[width=\linewidth]{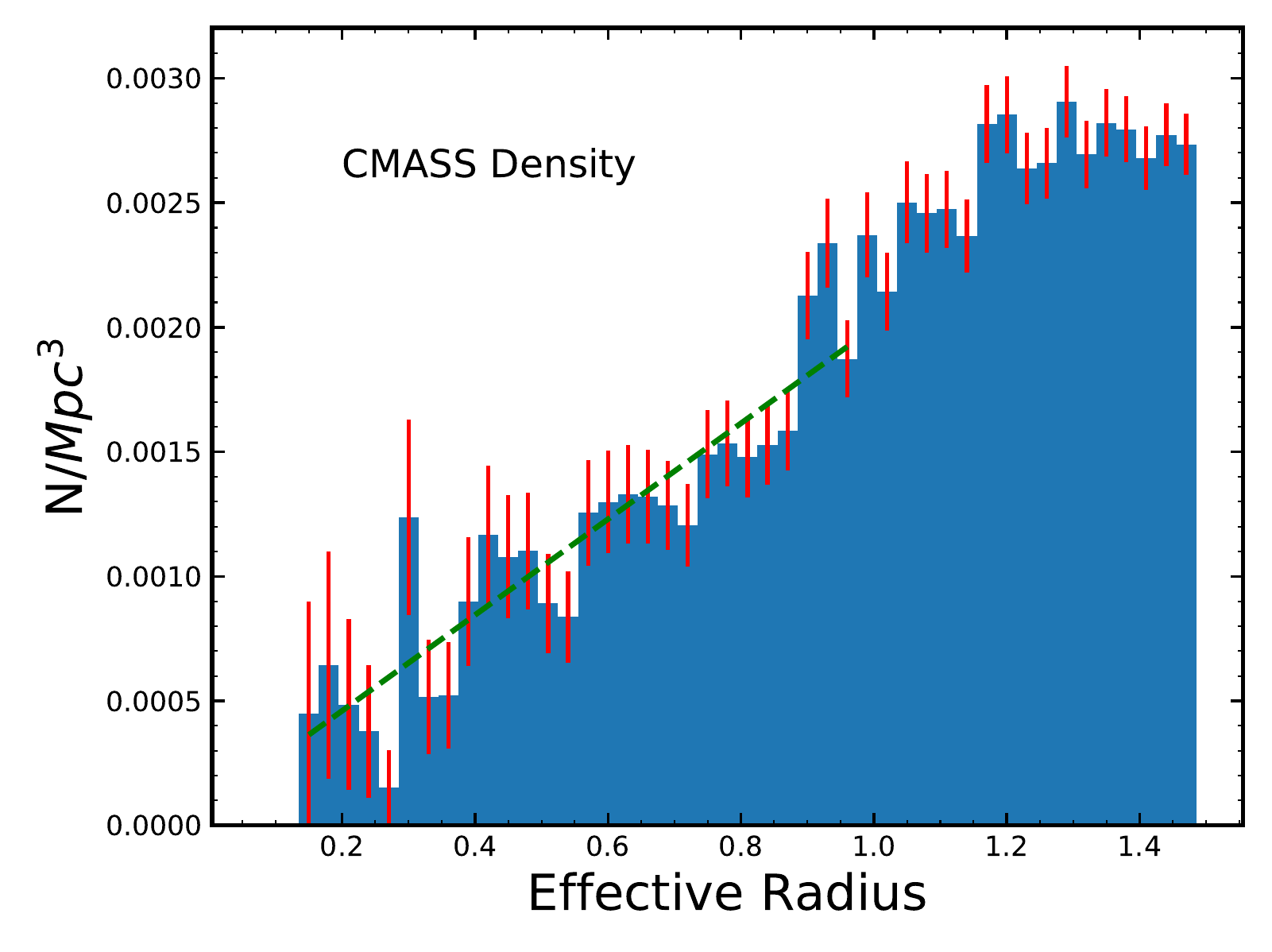}\par 
    \includegraphics[width=\linewidth]{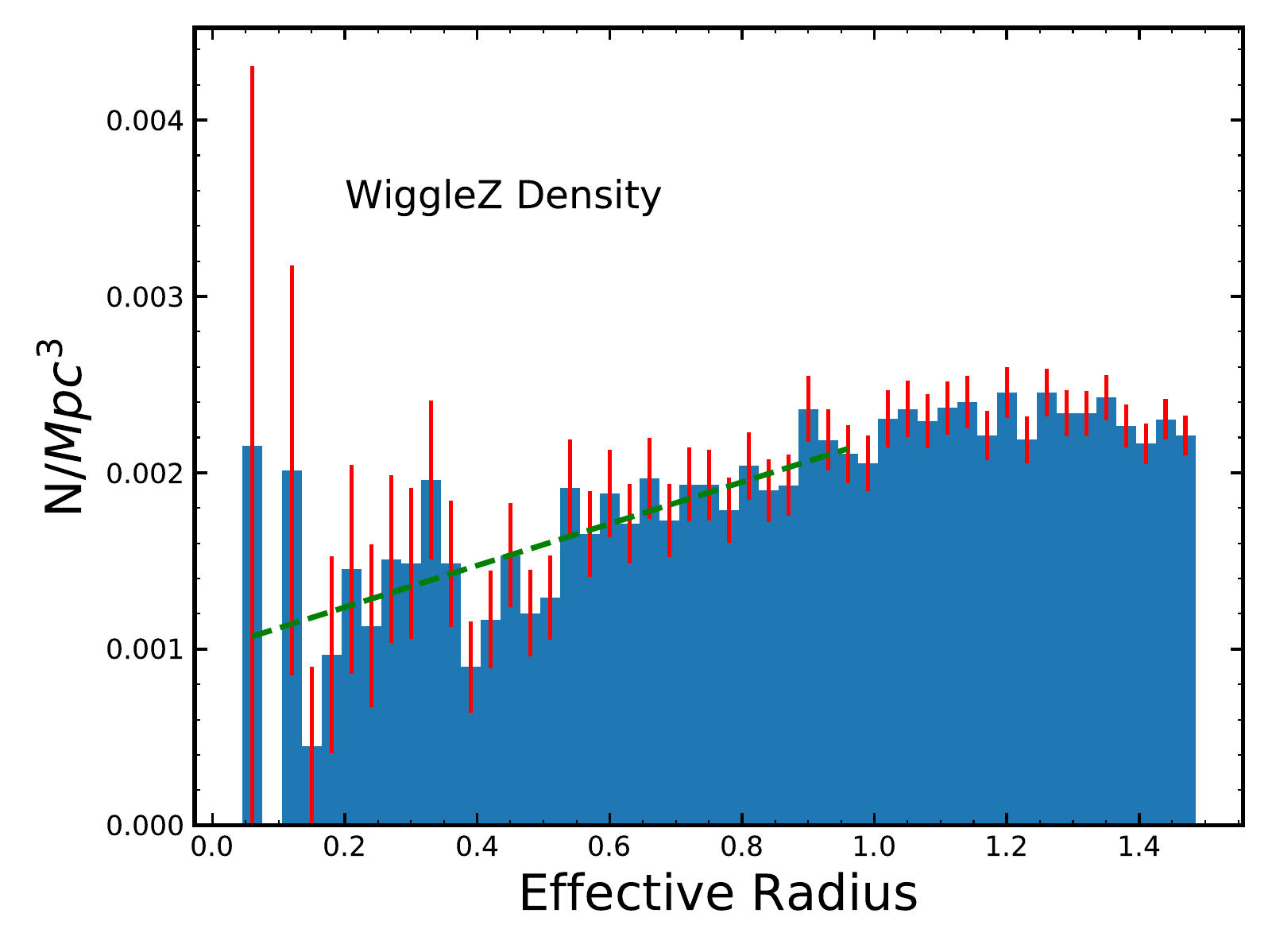}\par 
    \end{multicols}
\caption{Galaxy number density as a function of void radius in $R_{eff}$ with the uncertainties shown in red. The left plot shows the CMASS sample, and the right shows the WiggleZ sample. The void signal is more apparent in these plots, with the density clearly increasing as we move out from the center of the void in both. Shown by the green dashed line, we fit both density profiles with  equation \ref{eq:4} up to $R_{eff} = 1$. The CMASS fit deviates from zero by 10$\sigma$, while the WiggleZ plot deviates from zero by 4.2$\sigma$. The weaker void signal in the WiggleZ galaxies may be attributable to the fact that the void catalog was constructed from CMASS galaxies. The plot shows that our catalog of galaxies truly contains void galaxies and the surrounding shell of wall galaxies.} \label{fig:5}
\end{figure*}

To check for the void signal further, it is more enlightening to examine the stacked density in units of effective radius, because this accounts for the various sizes of the voids. This provides a cleaner look at the distribution of the galaxies across all voids. In this case, we expect the galaxy density to increase as we move out from a void $R_{eff}$. Figure \ref{fig:5} shows the stacked galaxy density vs distance from the void centers in $R_{eff}$ for both the CMASS and WiggleZ samples. The errors are once again calculated with equation \ref{eq:2}, resulting in some large error bars near the void centre. We quantify the plots with a linear fit out to $R_{eff} = 1$, given by 
\begin{equation}\label{eq:4}
Ratio = 10^{-3} C \frac{r}{R_{eff}} + D.  
\end{equation} 
The CMASS plot has a slope of 1.92 $\pm$ 0.18, giving a 10$\sigma$ deviation from zero. The WiggleZ plot has a slope of 1.18 $\pm$ 0.28, giving a 4.2$\sigma$ deviation from zero. Here, the void profiles are significantly detected in both the CMASS and WiggleZ samples.

\begin{figure*}
\begin{multicols}{2}
    \includegraphics[width=\linewidth]{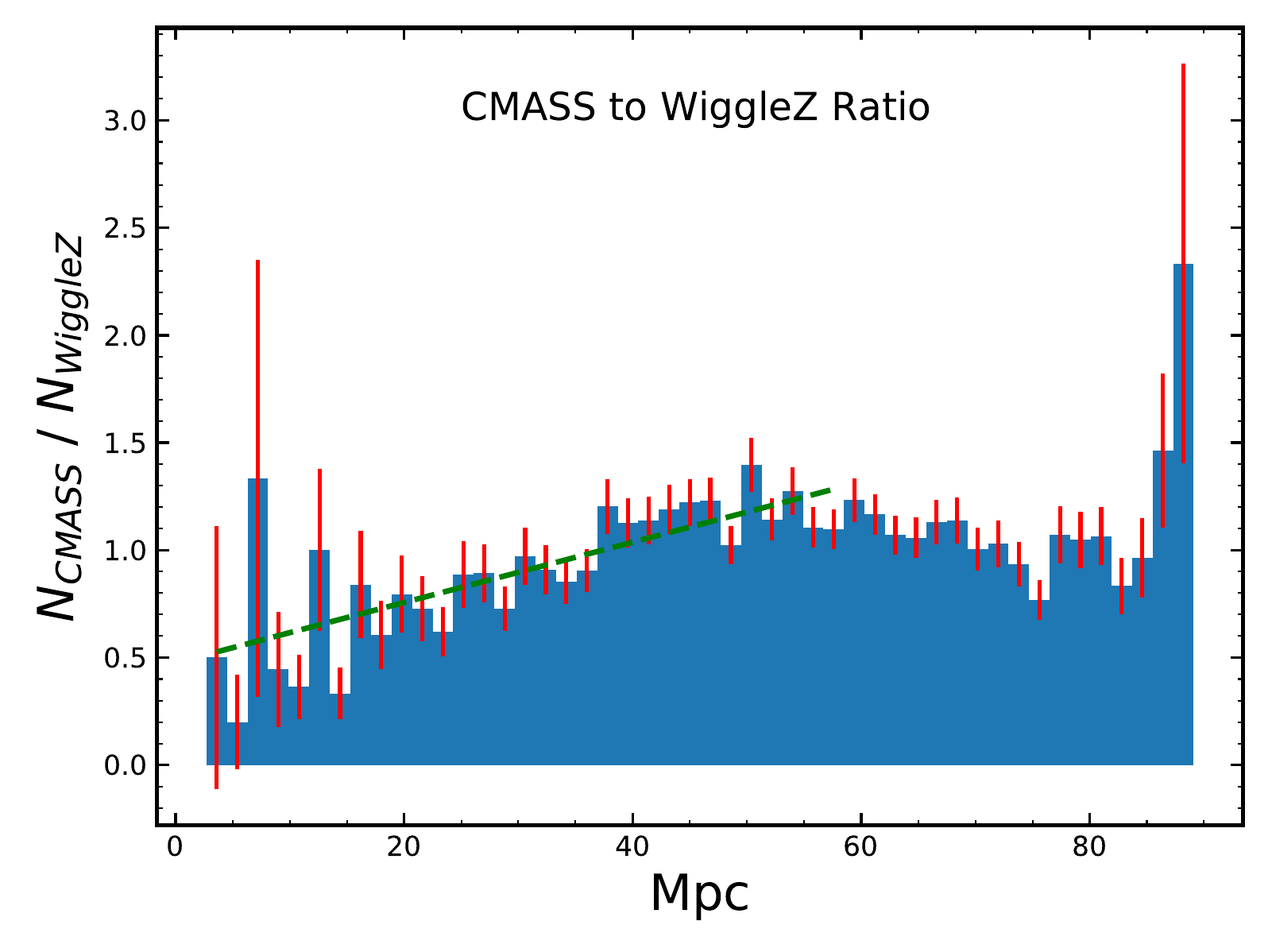}\par 
    \includegraphics[width=\linewidth]{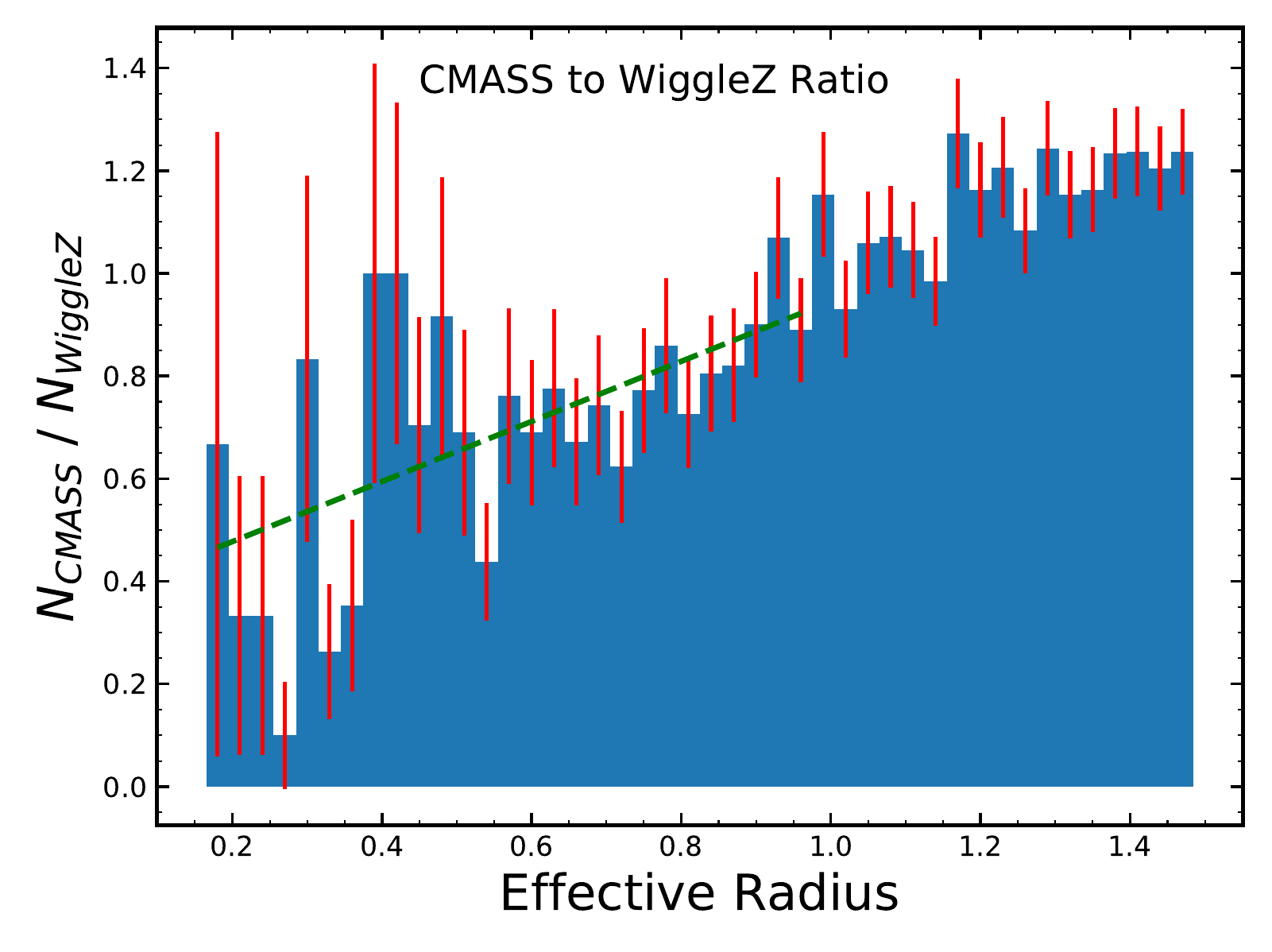}\par 
    \end{multicols}
\caption{The number ratio of CMASS (red) to WiggleZ (blue) galaxies in the cosmic voids. We plot the ratio vs distance from void centers in both the physical (Mpc) and $R_{eff}$ units. There is clearly a positive correlation between the ratio and the distance from the void centers. The linear fits are shown in green lines and the fit range for the Mpc and $R_{eff}$ are up to 60 Mpc and $R_{eff} = 1$, respectively. The Mpc plot has a slope of 701$\pm$121 while the $R_{eff}$ plot has a slope of 585$\pm$168, yielding 5.8$\sigma$ and 3.5$\sigma$ deviations from the zero slope, respectively. This correlation implies a deficit of red galaxies in voids compared to walls.}\label{fig:6}
\end{figure*}

Figure \ref{fig:5} particularly illustrates that we can be confident that the void catalog we are using has truly identified under-dense regions of the galaxy density field. However, it is important to keep in mind that the WiggleZ population has a slightly weaker void signal, interpreted as the slope's deviation from zero. This may be attributable to the fact that the void positions are derived from the SDSS sample by \citet{mao2017}. We may now proceed with comparing properties of void and wall galaxies; of particular interest for our analysis are the red-to-blue galaxy number ratio and color differences between void and wall galaxies. First, we plot the ratio of $N_{CMASS}/N_{WiggleZ}$ galaxies in both the physical and $R_{eff}$ units in Figure \ref{fig:6}. While it is possible that the normalization of the ratios are biased due to survey incompleteness and the exact value depends on the magnitudes and other selections, the observed slope is less subjective to the selection biases.
Figure \ref{fig:6} shows that the red-to-blue galaxy ratio increases with increasing distance from the void center. Using equation \ref{eq:2}, the ratio vs Mpc plot has a slope of 701 $\pm$ 121 and thus a 5.8$\sigma$ deviation from zero. Using equation \ref{eq:4} the ratio vs $R_{eff}$ has a slope of 585 $\pm$ 168 and a 3.5$\sigma$ deviation from zero. Thus, the increase in the ratio with increasing distance from void center is significant, implying that voids have a larger blue galaxy population percentage than walls. This is a result which has been found before \citep{szomoru1996,popescu1997,grogin1999, rojas2003, rojas2004, sorrentino2006, tavasoli2015}, though here we use two distinct galaxy surveys.

Our large sample size also offers us the ability to investigate whether void galaxies display color differences when compared to wall galaxies is due to a true color difference between their respective blue and red galaxies. 
Again, we define galaxies within one $R_{eff}$ as void galaxies and those within 1 $R_{eff}$ < d < 1.5 $R_{eff}$ as wall galaxies.
We employ the Kolmogorov--Smirnov (KS) test to determine the likelihood that two samples came from the same parent population. We begin with testing the combined population of CMASS and WiggleZ galaxies. In particular, we test the $g - r$ colors of void and wall galaxies from both samples. Such a KS test returns a KS-statistic of 0.092 and a p-value of $6.9\times10^{-18}$, demonstrating that the wall and void galaxies are sampled from different parent populations when combining the CMASS and WiggleZ samples.
We find a mean $g - r$ color of 1.298 $\pm$ $0.001$ for wall galaxies and 1.210 $\pm$ $0.002$ for void galaxies. The uncertainties of the mean color for a population are given by 
\begin{equation}\label{eq:5}
\delta = \frac{\overline{\sqrt{\delta_{M_g}^2 + \delta_{M_r}^2}}}{\sqrt{N}},  
\end{equation} 
where $N$ is the population size, and $\delta_{M_g}$ and $\delta_{M_r}$ are the magnitude errors used in the color calculation. We adopt the median uncertainties for SDSS CMASS galaxies in 0.43<z<0.70 of 0.04 for the $r$ band and 0.1 for the $g$ band for each galaxy. 
This simple mean comparison shows that the void and wall galaxies have different mean colors when combining the CMASS and WiggleZ samples.
Since the student's t-test is a more robust statistical test to compare the means of two samples, we apply the test to the color distributions of wall and void galaxies, yielding a t-statistic of 7.9 and a p-value of $3.3\times10^{-15}$. Evidently, the apparent color difference between the void and wall populations is very significant, confirming the KS test result.
Since we are combining the CMASS and WiggleZ samples in the above analysis, the color difference between the void and wall galaxies could be explained by either the voids having a larger percentage of blue galaxies than walls, or the void and wall galaxies having a true intrinsic color difference. 

Next, we perform the KS and student's t-tests between void and wall galaxies in either the CMASS or WiggleZ sample, separately.  This is a more likely analysis for detecting any true color evolution between void and wall galaxies, because this is testing in a much smaller color range and largely removes the color changing effect from a pure density evolution.
The KS test yields a KS-statistic of 0.030 and a p-value of 0.21 for $g - r$ color between void and wall galaxies for the WiggleZ population. 
For the CMASS sample, the corresponding numbers are 0.016 for the KS-statistic and 0.93 for the p-value.
Thus, we cannot reject the null hypothesis that the color of the both blue galaxies and red galaxies across the void and wall populations do not significantly differ from each other. 
The WiggleZ wall galaxies have a mean $g - r$ color of 0.800 $\pm$ $0.002$, while the WiggleZ void galaxies have a $g - r$ color of 0.793 $\pm$ $0.003$. The uncertainties of these mean colors are once again given by Equation \ref{eq:5}. A student's t-test of the $g - r$ color in the WiggleZ sample yields a t-statistic of 0.78 and a p-value of 0.44. Again, we cannot conclude that blue galaxies in either the wall or void populations have different mean colors.
The CMASS wall galaxies have a $g - r$ color of 1.729 $\pm$ $0.001$ and the CMASS void galaxies have a $g - r$ color of 1.731 $\pm$ $0.003$. A student's t-test of the $g - r$ color in the CMASS sample yields a t-statistic of -0.24 and a p-value of 0.81. Similar to what we have just seen, we cannot conclude that red galaxies in either the wall or void populations are significantly redder. All of our statistical test results are outlined in Table 1.

Combining these results with the increasing ratios demonstrated in Figure \ref{fig:6} shows that the result of measuring an apparent average bluer color in the void population is the result of varying population densities for red and blue galaxies, but not a true color difference. 
Since our samples have been pre-selected to only include luminous and normal galaxies, our analysis results only apply to those galaxies as well.
It should also be noted that these results were found to be dependent on the initial void sample, in particular the probability threshold imposed to make sure the voids are not the result of Poisson noise. When the probability threshold is relaxed (more voids but less certain of their validity), we find that blue galaxies in voids are significantly bluer than blue galaxies in walls.  Since this result is based on a less pure void sample, we will not discuss its implications further.
We find that our results are robust against nearly all other sample cuts we have made, including our choice for the minimum void radius, absolute magnitude bright and faint cutoffs for both the CMASS and WiggleZ samples, color cut for the CMASS sample, histogram bin sizes, and survey boundaries.

\begin{table*}
 \caption{A comparison of wall and void galaxy color statistics using the CMASS and WiggleZ samples individually or jointly.}
 \label{tab:example}
 \begin{tabular}{ccccccc}
  \hline
  Wall vs Void Population & Wall Mean $g - r$ & Void Mean $g - r$ & KS P-Value & KS-Value & T-Test P-Value & T-Value\\
  \hline
  CMASS $g - r$ & 1.729 & 1.731 & 0.93 & 0.016 & 0.81 & $-0.24$\\ 
  WiggleZ $g - r$ & 0.80 & 0.79 & 0.21 & 0.030 & 0.44 & 0.78 \\
  Combined $g - r$ & 1.298 & 1.210 & $6.9\times10^{-18}$ & 0.092 & $3.3\times10^{-15}$ & 7.9 \\
  \hline
 \end{tabular}
\end{table*}

\section{Comparing Voids of Various Sizes}
We also split the void catalog into three bins of varying sizes to investigate whether the red-to-blue galaxy number ratio depends on the void effective radius. In this analysis, we use all voids with effective radius between 25 Mpc and 100 Mpc. We then split these voids into three bins: $25 < R_{eff} < 50$, $50 < R_{eff} < 75$, and $75 < R_{eff} < 100$, which we will refer to as the small, medium, and large voids, respectively. The small void bin contains 128 voids with 4372 CMASS galaxies and 3362 WiggleZ galaxies. The medium void bin contains 64 voids with 4017 CMASS galaxies and 4638 WiggleZ galaxies. The large void bin contains 9 voids with 696 CMASS galaxies and 1082 WiggleZ galaxies. 

To test whether larger voids exhibit a larger amount of deficit of red galaxies than smaller voids, we compare the red-to-blue galaxy ratios for each of the three bin sizes. Figure~8 shows the three void bins densities as a function of the Mpc radius, while Figure \ref{fig:9} uses effective radius. With such few voids and relatively few galaxies, the 75 to 100 Mpc bin does not yield much useful insight. We once again fit the ratios with linear fits to characterize them --- the Mpc plots are fit with equation \ref{eq:2} and the $R_{eff}$ plots are fit with equation \ref{eq:4}. The Mpc plots seem to show evidence that the small voids actually have a larger deficit of red galaxies than the medium voids, opposite of what one might suspect. It would seem natural for larger voids to be more sparse than smaller voids. The small voids ratio vs Mpc fit has a slope of 1143 $\pm$ 262 and the medium voids a slope of 285 $\pm$ 135. This yields a 4.4$\sigma$ and 2.1$\sigma$ deviation from zero, respectively. The small voids ratio vs $R_{eff}$ fit has a slope of 612 $\pm$ 286 and the medium voids a slope of 427 $\pm$ 252, yielding a 2.1$\sigma$ deviation from zero for the small voids and a 1.7$\sigma$ deviation from zero for the medium voids.

\begin{figure*}
\begin{multicols}{3}
    \includegraphics[width=\linewidth]{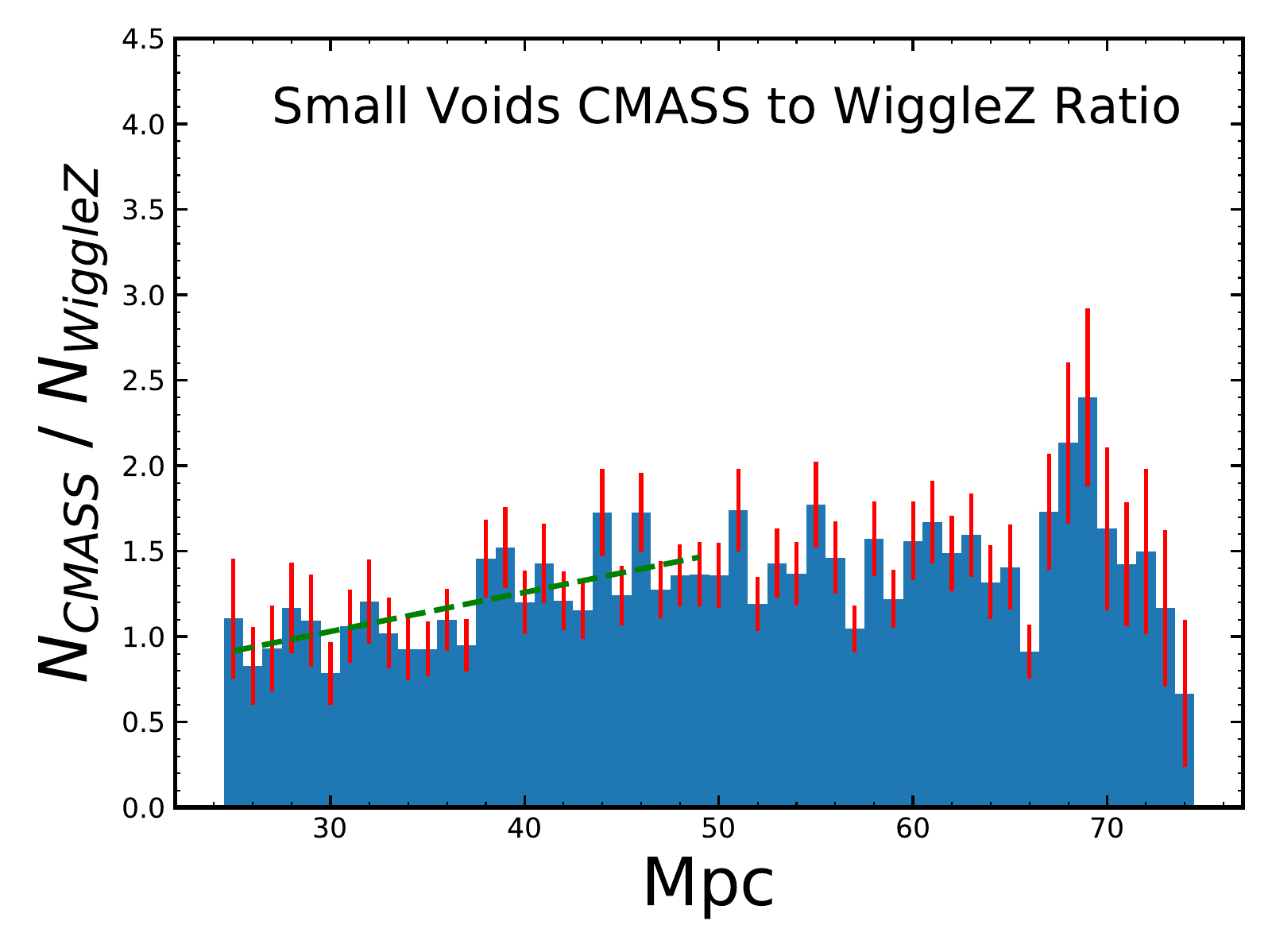}\par 
    \includegraphics[width=\linewidth]{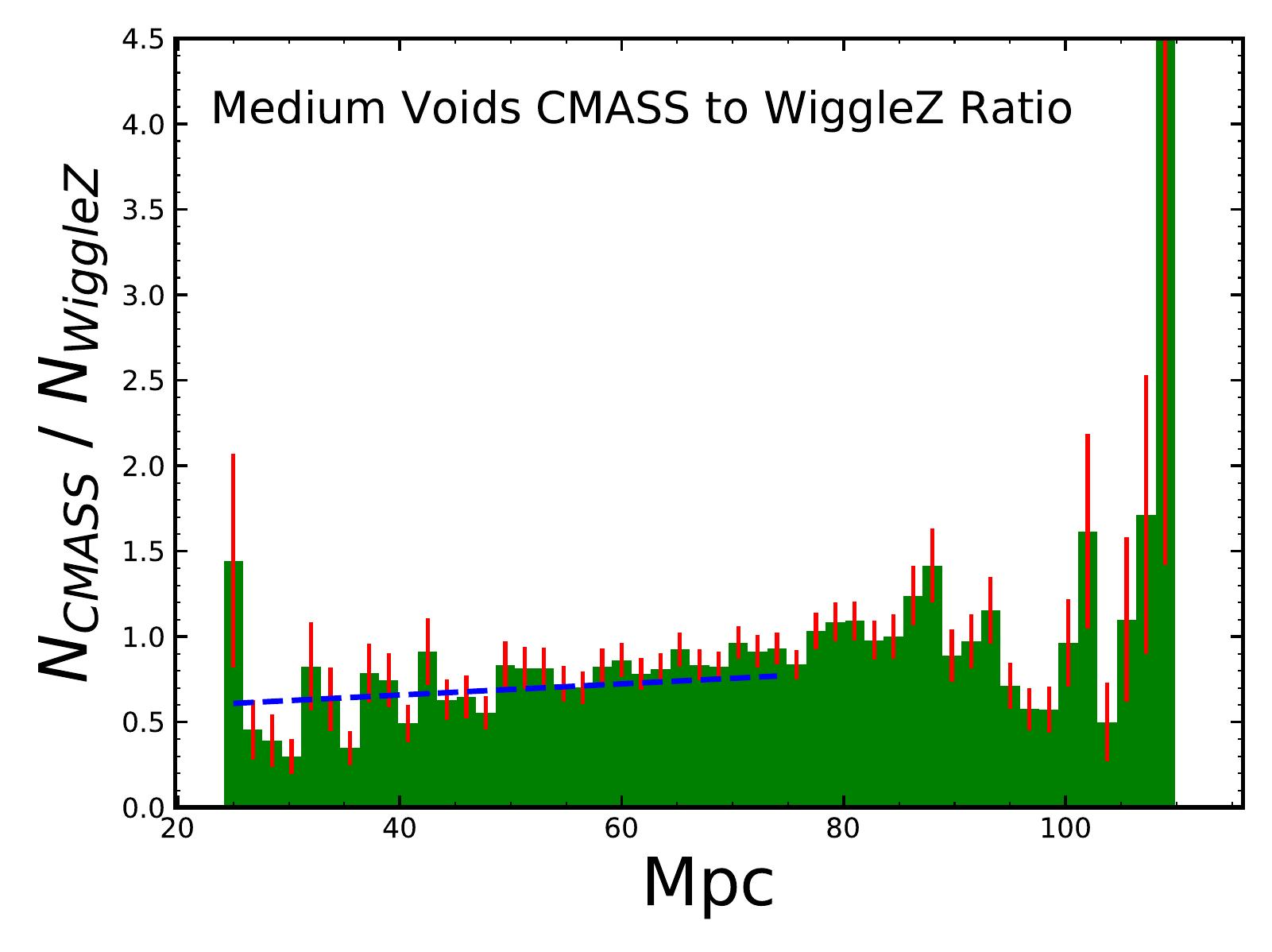}\par 
	\includegraphics[width=\linewidth]{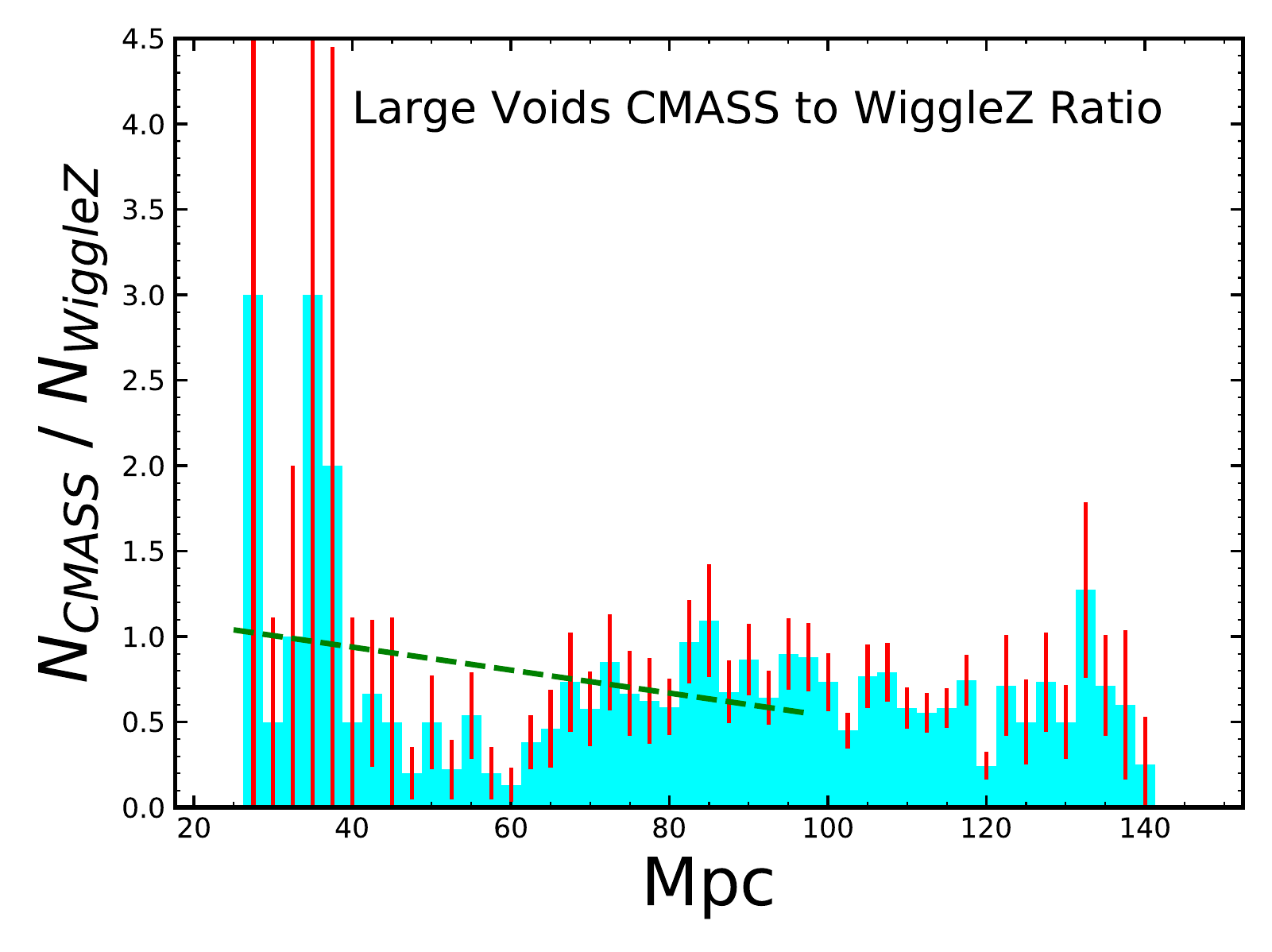}\par 
    \end{multicols}
\caption{CMASS-to-WiggleZ galaxy ratio plots as functions of distance from void center in Mpc for small, medium, and large voids respectively from the left to right.  We fit all the plots with the linear equation \ref{eq:2} to test if the slope is dependent on void size. The small voids' fit has a slope of 1143 $\pm$ 262, 4.4$\sigma$ deviation from the zero slope, and the medium voids' fit has a slope of 285 $\pm$ 135, 2.1$\sigma$ deviation from zero. The slopes of small voids and medium voids differ by $2.9\sigma$. The large voids' ratio profile is dominated by measurement uncertainties.}
\end{figure*}\label{fig:8}

\begin{figure*}
\begin{multicols}{3}
    \includegraphics[width=\linewidth]{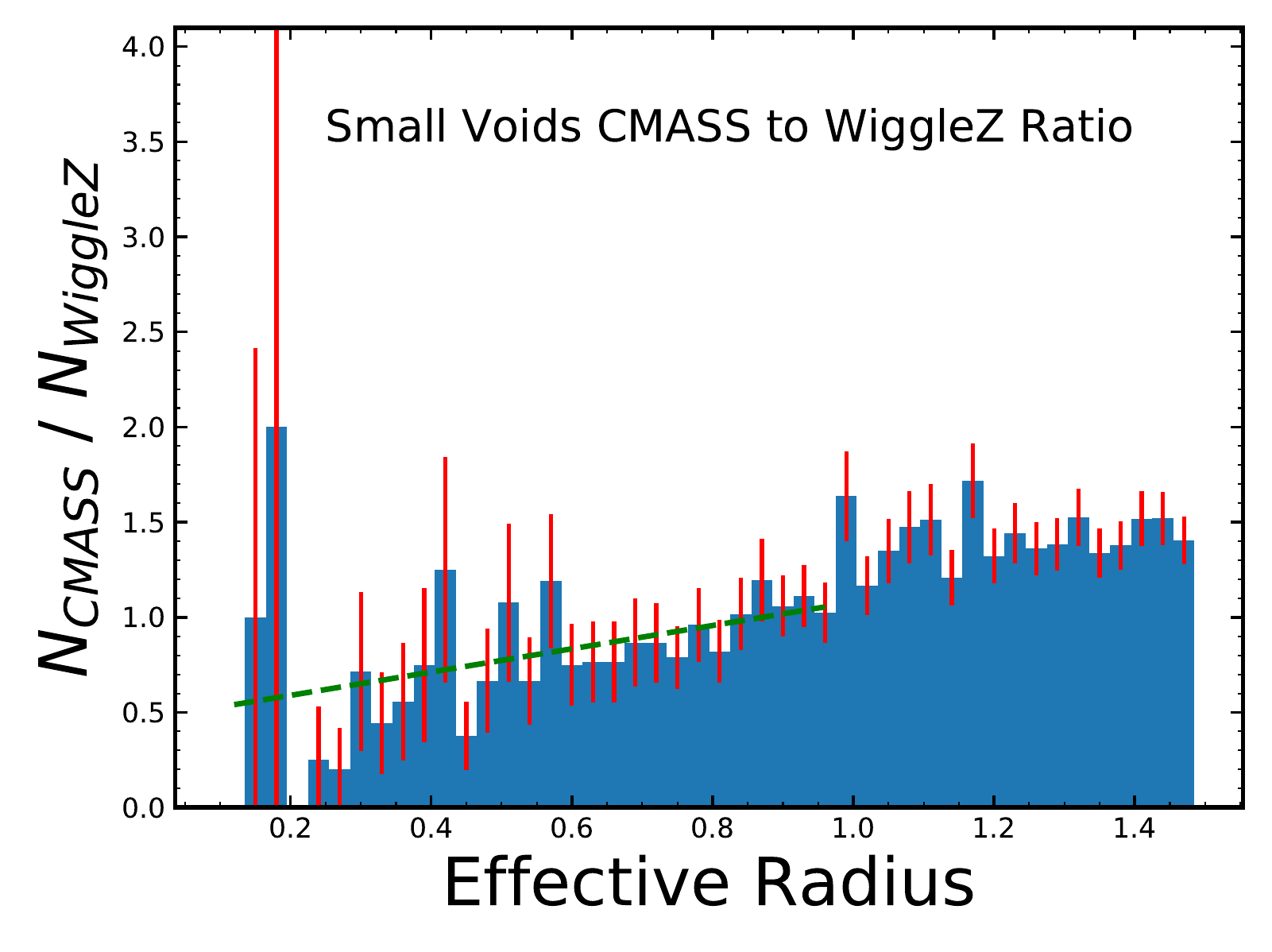}\par 
    \includegraphics[width=\linewidth]{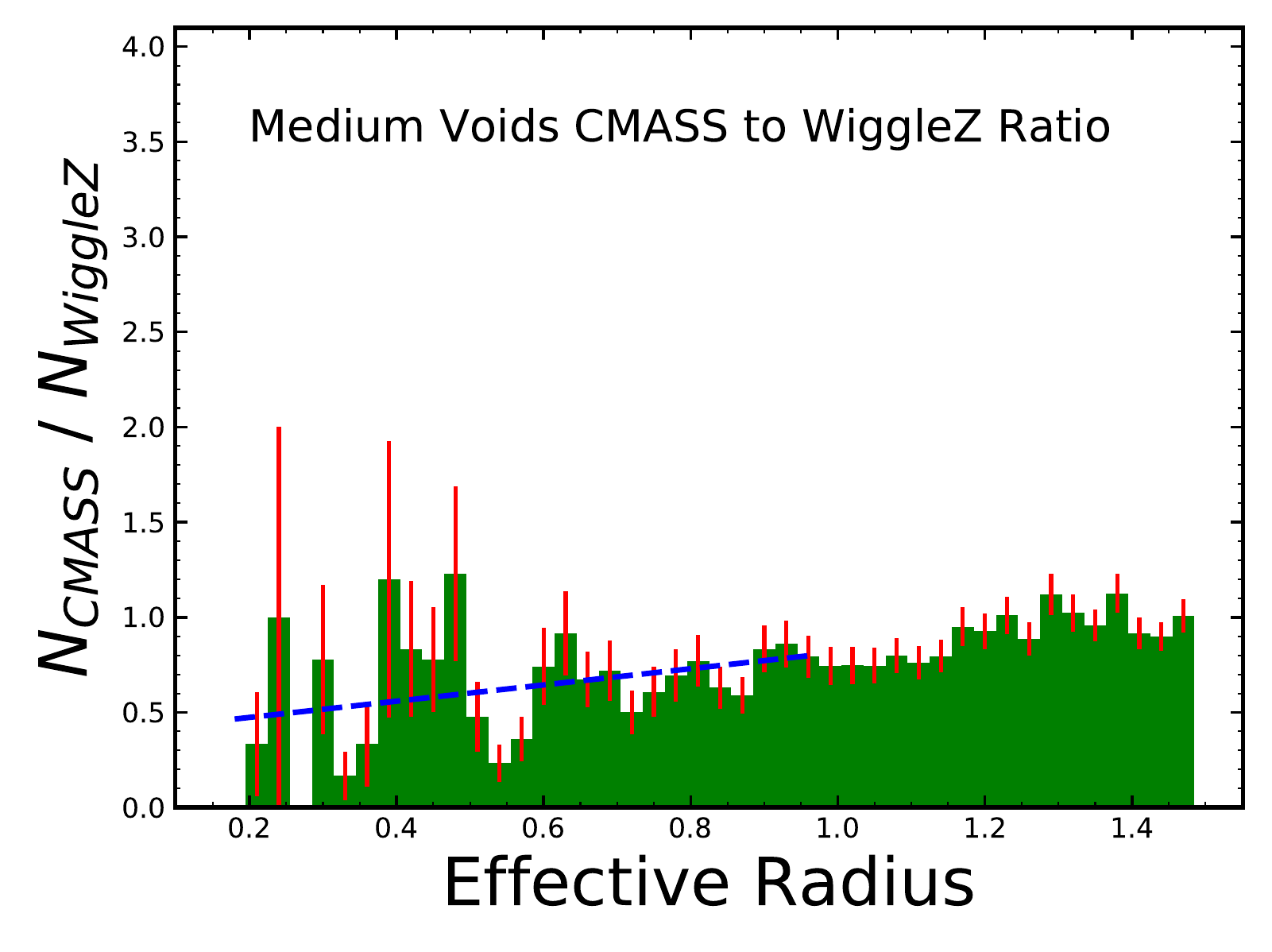}\par 
	\includegraphics[width=\linewidth]{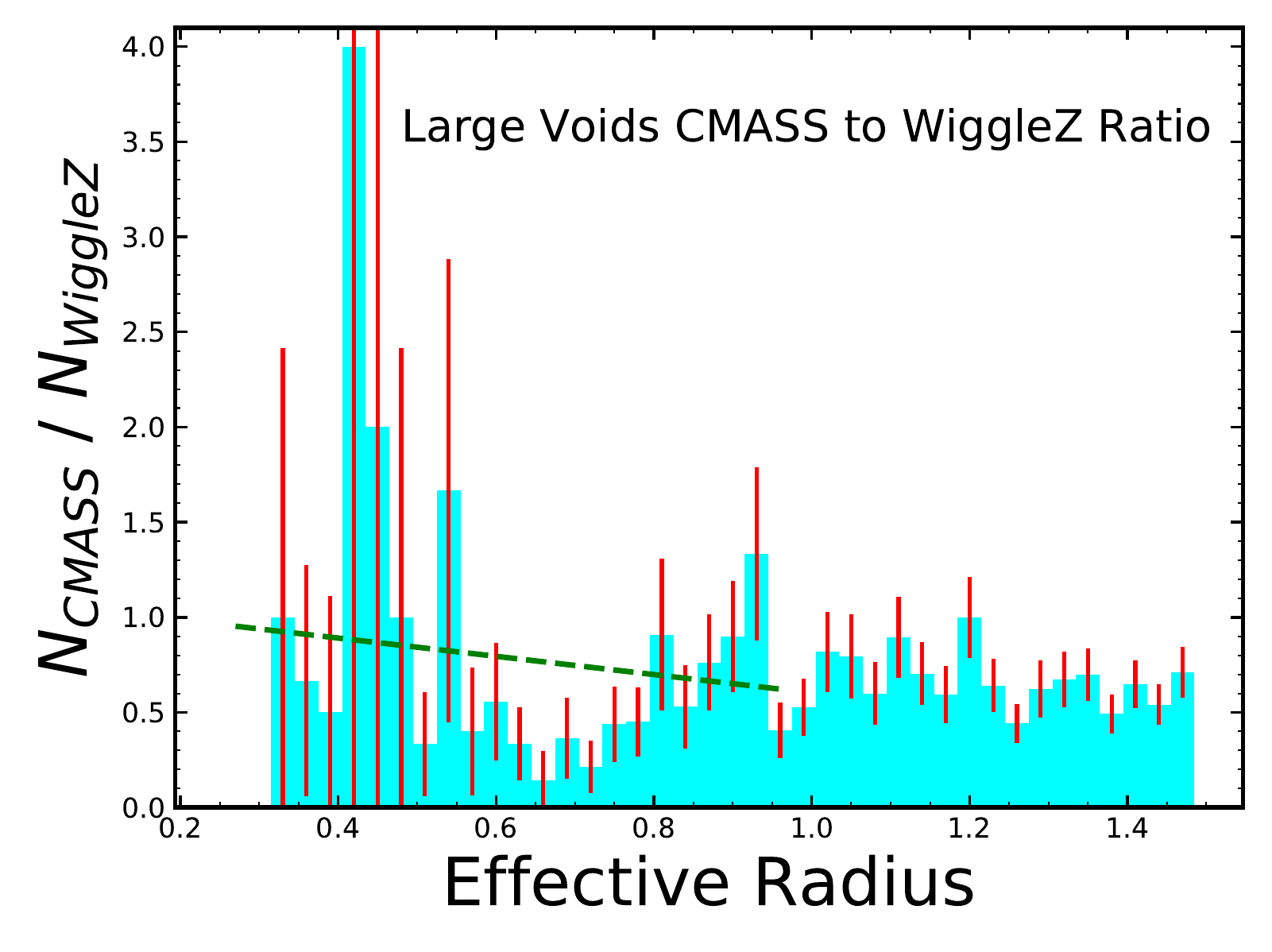}\par 
    \end{multicols}
\caption{MASS-to-WiggleZ galaxy number ratio plots as functions of distance from the void center in $R_{eff}$ for small, medium, and large voids respectively from the left to right.
The small voids fit has a slope of 612$\pm$286 with a 2.1$\sigma$ deviation from zero, while the medium voids fit has a slope of 427$\pm$252 and deviates from zero by 1.7$\sigma$. We cannot conclude that there is a significant difference between the small and medium voids red-to-blue ratio profiles. Once again, the large voids ratio profile is dominated by measurement uncertainties.}\label{fig:9}
\end{figure*}
\subsection{Comparison to Theoretical Predictions}

We compare the observational results in Table 1 to results from a theoretical prediction utilizing the Dark SAGE semi-analytic model \citep{stevens2016}.  Dark SAGE treats the evolution of baryons, and hence galaxies, as a post-processing step of an
N-body simulation. While fine structural detail of galaxies is lost, orders of magnitude of computational efficiency is gained through the semi-analytic approach. The details of the Dark SAGE model can be found in \citet{stevens2016} and we run the model on the Millenium simulation \citep{springel2005}. For the sake of simplicity, we only
present DARK SAGE as run on one simulation and we chose
Millennium because the underlying SAGE codebase is well tested on
this simulation, the data are readily available, and this makes the
results of this paper more closely comparable to the multitude of
semi-analytic publications that have used this simulation.

In Figure~10 we demonstrate agreement between the prediction of the semi-analytic model with the results from our analysis.  In particular, we calculate SDSS $g-r$ colors for all halos in the simulation volume as a function of the number of neighbors within an 8\,Mpc sphere.  We then use the number of neighbors to distinguish between void and wall galaxies (denoted by dashed lines), where void galaxies have fewer than 25 neighbors in 8\,Mpc, and wall galaxies have between 50 and 400 neighbors.  The gold stars denote the mean color for void and wall galaxies, resulting in an average color difference of 0.1\,mags, in good agreement with the value determined in this paper (0.09\,mag; see Table 1). The green lines on Figure~10 are the binned data from the model, where each bin contains the same number of galaxies and the length of the line represents the range of neighbor counts included within the bin.  Figure~10 also clearly demonstrates that void galaxies are bluer than their counterparts in more dense environments. 
\begin{figure*}
%\begin{multicols}{3}
    \includegraphics[width=0.5\linewidth]{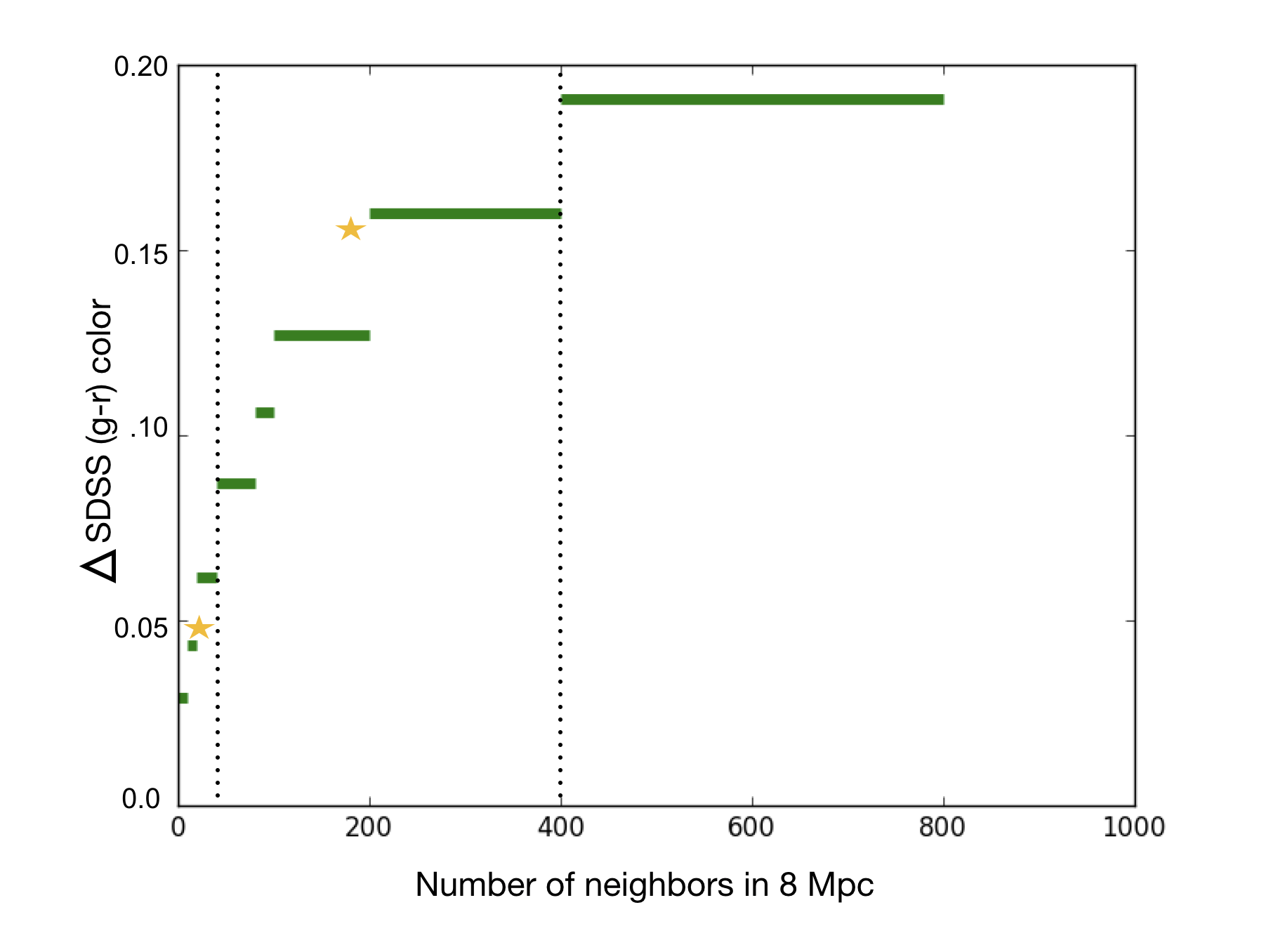}\par
%    \end{multicols}
\caption{Void and wall galaxy color differences from a semi-analytic simulation model, where void galaxies (left to the first dotted line) and wall galaxies (between the two dotted lines) are defined based on their number of neighbors within 8\,Mpc in the simulation.  Stars show the average colors for the void and wall galaxies, and the average color difference of 0.1 in the model matches the observed value 0.09 (Table~\ref{tab:example}).   
}\label{fig:10}
\end{figure*}

\section{Discussion and Conclusions}
We construct a catalog of void galaxies along with nearby wall galaxies to probe galaxy densities and color differences between the two populations. Our sample is selected from a publicly available void catalog produced by \citet{mao2017}, BOSS CMASS data, and the WiggleZ Dark Energy Survey. After identifying void galaxies through their accurate distances from spectroscopic redshifts, we test for the void signal to demonstrate that our catalog galaxies lie in and around voids. We then analyze the number ratio between CMASS and WiggleZ galaxies as a function of distance from the void centers, demonstrating that voids tend to have proportionally fewer red galaxies than walls galaxies.
Finally, we use KS and student's T tests to compare galaxy colors between void and wall galaxies either using the CMASS and WiggleZ samples jointly or individually.  While significant color difference is found between void and wall galaxies in the CMASS-WiggleZ joint sample, we find no significant color difference between the void and wall galaxies when analyzing the CMASS or WiggleZ samples individually.
This shows that void galaxies do not tend to be truly bluer than wall galaxies at our magnitude range of $M_R \simlt -20$, indicating that true color difference between wall and void galaxies does not contribute to the observed average color difference when combining the red and blue galaxy samples. This, combined with our red-to-blue galaxy number ratio analysis, indicates that a deficit of red galaxies, and not a true color difference, is the driver of the observed color difference between void and wall galaxies. 

These results have implications for how the environment affects the evolution of galaxies --- the deficit of red galaxies in the void environment can be the results of less frequent mergers in voids compared to high density regions.  However, galaxies assembled in voids, while more likely to be late-type blue galaxies, may follow evolution paths similar to their counterparts in walls, since we do not find significant intrinsic color difference between the void and wall galaxy samples for either red or blue galaxies to absolute magnitude limits of $\sim -20$ -- $-21$ for blue and red galaxies, respectively, which is close to the $M* \sim -21.8$ of galaxies at $z\sim0.5$ \citep[e.g.,][]{bell2003, dai2009}.  Thus, our analysis results apply to luminous and normal galaxies.

There are several previous analyses on the topic of color difference between void and wall galaxies.  While studies of nearby voids have the advantage of probing the low-end of the luminosity function to dwarf galaxies, these analyses do not have large enough sample volume to build up a large sample of luminous void galaxies.  Our analysis, in contrast, has the largest sample of luminous and normal galaxies, and has no sensitivity for dwarf galaxies.  Ideally, we should combine the advantage from both analyses from nearby or higher redshift void studies to comprehensively understand void galaxies.

\citet{rojas2004} showed that there is an observable $g - r$ color difference between nearby ($comoving-distance < 260~Mpc$) void and wall galaxies in the luminosity range of $-22.5 < M_r < -13.5$ with a KS p-value of 0.002, while having no statistically significant difference in their Sersic indices. Further, their void and wall galaxy samples have similar surface profiles, indicating that the color difference may have some ground in true color difference, in contrast to our analysis results.  When splitting the samples by Sersic indices at $n = 1.8$, as a proxy for splitting by galaxy type, they found void galaxies were significantly bluer in both of the child populations, even when accounting for magnitude differences.
\citet{hoyle2012} performed much of the same analysis as \citet{rojas2004}, though with the advantage of using the later data releases of SDSS, increasing population size of both the voids and void galaxies. They largely find results similar to \citet{rojas2004}, though with more statistical significance. Again, the sample studied by \citet{hoyle2012}, with a max redshift of $z_{max} = 0.107$, is still significantly less distant than the void galaxies we study in this analysis. This allows the authors to study much more faint galaxies than we have access to, $-22.5 < M_r < -13.5$ for \citet{rojas2004} and $-23 < M_r < -12$ for \citet{hoyle2012}.
It is worth noting that our lack of color difference is consistent with the findings in \citet{penny2015}, wherein the authors found no significant color difference between void and non-void galaxies sub-classified as strong AGN, weak AGN, star forming, retired, and passive galaxies. These results were found with galaxies up to $z\sim0.1$ and $M_r < -18.4$. The key difference then, between our study, \citet{penny2015}, and those of \citet{rojas2004} and \citet{hoyle2012}, seems to be the magnitude depth achieved, though the lack of color difference between our blue void and wall galaxies is still in tension with \citet{hoyle2012}'s findings of a color difference at all luminosities, possibly due to their relatively small sample size for normal and luminous galaxies.

Combining the results from this study, which has the largest sample for the normal and luminous galaxies, and those from local voids, a converging picture is that there is little environmental-dependence galaxy evolution for normal and luminous galaxies between the void and wall environments, except that the red galaxy numbers are dictated by the mergers which is environmental-dependent.
For dwarf galaxies, the environmental effects are much stronger and can result in intrinsic color differences for dwarf galaxies in the void or wall regions.

It is of importance to emphasize that in this study, the void catalog is produced using CMASS data. This may well introduce a selection bias between the red and blue populations. It would be informative to reproduce these results with a void catalog produced from a blue galaxy survey, as this study, as well as most other void studies, derive their voids from surveys which favor red galaxies, particularly due to SDSS's large sample size. Future surveys, or even the full WiggleZ catalog, could make such analysis possible.

\section*{Acknowledgements}

We thank J.\ N.\ Bregman, C.-H.\ Chuang, and Y.\ Wang for the helpful discussion and comments.
We acknowledge the financial support from the NASA ADAP programs NNX15AF04G, NNX17AF26G, NSF grant AST-1413056.

%%%%%%%%%%%%%%%%%%%%%%%%%%%%%%%%%%%%%%%%%%%%%%%%%%



%\bibliographystyle{mnras}


% Don't change these lines
\bsp	% typesetting comment
\label{lastpage}
\end{document}